

\font\fiverm=cmr5
\font\fivei=cmmi5       \skewchar\fivei='177
\font\fivesy=cmsy5      \skewchar\fivesy='60
\font\sixrm=cmr5 at 6pt
\font\sixi=cmmi5 at 6pt \skewchar\sixi='177
\font\sixsy=cmsy5 at 6pt        \skewchar\sixsy='60
\font\sevenrm=cmr7
\font\seveni=cmmi7                      \skewchar\seveni='177
\font\sevensy=cmsy7                     \skewchar\sevensy='60
\font\eightrm=cmr7 at 8pt
\font\eighti=cmmi7 at 8pt       \skewchar\eighti='177
\font\eightsy=cmsy7 at 8pt      \skewchar\eightsy='60
\font\tenrm=cmr10
\font\tenbf=cmbx10
\font\tenit=cmti10
\font\teni=cmmi10               \skewchar\teni='177
\font\tensy=cmsy10              \skewchar\tensy='60
\font\tenex=cmex10
\font\tensl=cmsl10
\font\elevenrm=cmr10 at 11pt
\font\elevenbf=cmbx10 at 11pt
\font\elevenit=cmti10 at 11pt
\font\eleveni=cmmi10 at 11pt            \skewchar\eleveni='177
\font\elevensy=cmsy10 at 11pt           \skewchar\elevensy='60
\font\elevensl=cmsl10 at 11pt
 \font\sixteenbf=cmbx10 at 16pt
\catcode`\@=11
\let\sectf=\elevenbf
\def\@setstrut{\setbox\strutbox=\hbox{%
  \vrule height .7\baselineskip depth .3\baselineskip width 0pt\relax}}
\def\@setmathskips{%
  \abovedisplayskip=0.50\baselineskip plus 0.25\baselineskip minus 1pt%
  \belowdisplayskip=\abovedisplayskip%
  \abovedisplayshortskip=0.25\baselineskip plus 0.25\baselineskip%
  \belowdisplayshortskip=0.50\baselineskip plus 0.25\baselineskip minus 1pt}
\textfont3=\tenex
\scriptfont3=\tenex
\scriptscriptfont3=\tenex
\def\tenpoint{
  \def\rm{\fam0\tenrm}%
  \textfont0=\tenrm \scriptfont0=\sevenrm \scriptscriptfont0=\fiverm
  \textfont1=\teni  \scriptfont1=\seveni  \scriptscriptfont1=\fivei
  \textfont2=\tensy \scriptfont2=\sevensy \scriptscriptfont2=\fivesy
  \textfont\itfam=\tenit\def\it{\fam\itfam\tenit}%
  \textfont\bffam=\tenbf\def\bf{\fam\bffam\tenbf}%
  \textfont\slfam=\tensl\def\sl{\fam\slfam\tensl}%
  \jot=2.75pt\normalbaselineskip=13pt\normallineskiplimit=2pt%
  \normallineskip=1pt\normalbaselines\@setmathskips\@setstrut\rm}
\def\elevenpoint{
  \def\rm{\fam0\elevenrm}%
  \textfont0=\elevenrm \scriptfont0=\eightrm \scriptscriptfont0=\sixrm
  \textfont1=\eleveni  \scriptfont1=\eighti  \scriptscriptfont1=\sixi
  \textfont2=\elevensy \scriptfont2=\eightsy \scriptscriptfont2=\sixsy
  \textfont\itfam=\elevenit\def\it{\fam\itfam\elevenit}%
  \textfont\bffam=\elevenbf\def\bf{\fam\bffam\elevenbf}%
  \textfont\slfam=\elevensl\def\sl{\fam\slfam\elevensl}%
  \jot=3pt\normalbaselineskip=13pt plus .2pt\normallineskiplimit=2pt%
  \normallineskip=1pt\normalbaselines\@setmathskips\@setstrut\rm}
 \newdimen\whang
\def\bibitem#1{{\hangindent=1em\hangafter=1%
 \parskip=1pt plus .5pt minus .25pt%
  \rightskip=0pt plus .2\hsize%
  \noindent\frenchspacing\tenpoint\baselineskip=12pt\@setstrut #1\par}}
\def\jr#1|#2|#3|#4|#5|#6|{\bibitem{#1 ($#2$). #3, {\it #4\/} {\bf #5,} #6.}}
\def\vr#1|#2|#3|#4|#5|#6|#7|{%
  \bibitem{#1 ($#2$). #3, in {\it #4,\/} ed. #5 (#6, #7).}}
\def\br#1|#2|#3|#4|#5|{\bibitem{#1 ($#2$). {\it #3\/} (#4, #5).}}
\def\pr#1|#2|#3|{\bibitem{#1 ($#2$). #3.}}
\def\list#1{  \goodbreak \vskip 6pt plus 1.8pt minus .9pt%
  \nobreak  \begingroup  \setbox0=\hbox{#1)\kern 1ex}%
  \whang=\wd0  \leftskip=\whang }
\def\endlist{\par\endgroup\goodbreak\vskip 6pt plus 1.8pt minus .9pt\nobreak}
\def\item#1{\par\noindent\llap{\hbox to\whang{\hfil#1)\kern1ex}}}
\def\bullist{\goodbreak\vskip 6pt plus 1.8pt minus .9pt\nobreak%
  \begingroup \whang=\parindent \leftskip=\whang%
  \parindent=0pt  \let\item=\bullitem }
\def\bullitem{\par\noindent\llap{\hbox to\whang{\hfil$\bullet$\hfil}}}
\def\unlist{ \goodbreak \vskip 6pt plus 1.8pt minus .9pt \nobreak%
  \begingroup \whang=\parindent \parindent=0pt %
  \parskip=3pt plus 1.5pt minus .75pt \let\item=\unitem}
\def\unitem{\par\hangindent=1em}
\elevenpoint
\topskip=10pt
\advance\vsize by \topskip
\widowpenalty=10000
\clubpenalty=10000
\parskip=0pt
\parindent=1em
\nonfrenchspacing
\normalbottom
\catcode`\@=12
 
{\obeylines\gdef\
{}}
\footline={\hfil}
\def\\{\cr}
\overfullrule=0pt
 \newcount\subchapnum\subchapnum=0
 \def\section#1#2{\removelastskip\goodbreak%
  \vskip 13pt plus 3pt minus 1.5pt \nobreak\setbox0=\hbox{#1}%
\vbox{{\rightskip=1em plus .5\hsize\leftskip=\rightskip%
 \parfillskip=0pt\parindent=0pt\parskip=0pt%
 \sectf\baselineskip=13pt\ifdim\wd0>3pt #1\hskip 6pt\fi#2\par}}%
\nobreak\vskip 6pt plus 1.5pt minus .5pt\nobreak\noindent}
\outer\def\subchap#1{ \global\advance\subchapnum by 1
   \section{\toghead\the \subchapnum}{#1}}
 \def\(#1){(\call{#1})}
\def\call#1{{#1}}
\def\taghead#1{}
\def\e{{\,\rm e}}
\def\12{{1\over2}}

\def\ie{{\it i.e. }}

\def\twiddle{\lower.9ex\rlap{$\kern -.1em\scriptstyle\sim$}}
\def\bigtwiddle{\lower1.ex\rlap{$\sim$}}
\def\gtwid{\mathrel{\raise.3ex\hbox{$>$\kern -.75em\lower1ex\hbox{$\sim$}}}}
\def\ltwid{\mathrel{\raise.3ex\hbox{$<$\kern -.75em\lower1ex\hbox{$\sim$}}}}
\def\bra#1{\left\langle #1\right|}
\def\ket#1{\left| #1\right\rangle}
\def\vev#1{\left\langle #1\right\rangle}
 \def\Re{{\cal R \mskip-4mu \lower.1ex \hbox{\it e}}}
\def\Im{{\cal I \mskip-5mu \lower.1ex \hbox{\it m}}}

\def\bar{\overline}

\def\cA{{\cal A}}
\def\cR{{\cal R}}

\def\one{{\bf 1}}

\def\Zint{\ {{\bf Z} \kern -.40em {\bf Z}}}
\def\real{{\vrule height 1.5ex width 0.05em depth 0ex
 \kern-0.06em {\bf R}}}
\def\natural{\ {\vrule height 1.5ex width 0.05em depth 0ex
 \kern-0.06em {\bf N}}}
\def\complex{\ {\vrule height 1.4ex width 0.05em depth 0ex
\kern-0.29em {\bf C}}}
\def\rational{\ {\vrule height 1.4ex width 0.05em depth 0ex
\kern-0.29em {\bf Q}}}
\def\cp{\complex {\vrule height 1.5ex width 0.05em depth 0ex
\kern-0.06em {\bf P}}}

\catcode`@=11
\newcount\tagnumber\tagnumber=0
\def\write@qn#1{}
\def\writenew@qn#1{}
\def\w@rnqwrite#1{\write@qn{#1}\message{#1}}
\def\@rrwrite#1{\write@qn{#1}\errmessage{#1}}
\def\taghead#1{\gdef\t@ghead{#1}\global\tagnumber=0}
\def\t@ghead{}
\expandafter\def\csname @qnnum+15\endcsname
  {{\t@ghead\advance\tagnumber by -5\relax\number\tagnumber}}
\expandafter\def\csname @qnnum+14\endcsname
  {{\t@ghead\advance\tagnumber by -4\relax\number\tagnumber}}
\expandafter\def\csname @qnnum+13\endcsname
  {{\t@ghead\advance\tagnumber by -3\relax\number\tagnumber}}
\expandafter\def\csname @qnnum+12\endcsname
  {{\t@ghead\advance\tagnumber by -2\relax\number\tagnumber}}
\expandafter\def\csname @qnnum+11\endcsname
  {{\t@ghead\advance\tagnumber by -1\relax\number\tagnumber}}
\expandafter\def\csname @qnnum0\endcsname
  {\t@ghead\number\tagnumber}
\expandafter\def\csname @qnnum+1\endcsname
  {{\t@ghead\advance\tagnumber by 1\relax\number\tagnumber}}
\expandafter\def\csname @qnnum+2\endcsname
  {{\t@ghead\advance\tagnumber by 2\relax\number\tagnumber}}
\expandafter\def\csname @qnnum+3\endcsname
  {{\t@ghead\advance\tagnumber by 3\relax\number\tagnumber}}
\def\callall#1{\xdef#1##1{#1{\noexpand\call{##1}}}}
\def\call#1{\each@rg\callr@nge{#1}}
\def\each@rg#1#2{{\let\thecsname=#1\expandafter\first@rg#2,\end,}}
\def\first@rg#1,{\thecsname{#1}\apply@rg}
\def\apply@rg#1,{\ifx\end#1\let\next=\relax%
\else,\thecsname{#1}\let\next=\apply@rg\fi\next}
\def\callr@nge#1{\calldor@nge#1-\end -}
\def\callr@ngeat#1\end -{#1}
\def\calldor@nge#1-#2-{\ifx\end#2\@qneatspace#1 %
  \else\calll@@p{#1}{#2}\callr@ngeat\fi}
\def\calll@@p#1#2{\ifnum#1>#2{\@rrwrite{Equation range #1-#2\space is bad.}
\errhelp{If you call a series of equations by the notation M-N, then M and
N must be integers, and N must ge M.}}\else%
{\count0=#1\count1=#2\advance\count1 by1\relax\expandafter%
\@qncall\the\count0, \loop\advance\count0 by1\relax%
    \ifnum\count0<\count1,\expandafter\@qncall\the\count0,%
  \repeat}\fi}
\def\@qneatspace#1#2 {\@qncall#1#2,}
\def\@qncall#1,{\ifunc@lled{#1}{\def\next{#1}\ifx\next\empty\else
  \w@rnqwrite{  \noexpand\(>>#1<<) undefined  }
  >>#1<<\fi}\else\csname @qnnum#1\endcsname\fi}
\let\eqnono=\eqno
\def\eqno(#1){\tag#1}
\def\tag#1$${\eqnono(\displayt@g#1 )$$}
\def\aligntag#1\endaligntag
  $${\gdef\tag##1\\{&(##1 )\cr}\eqalignno{#1\\}$$
  \gdef\tag##1$${\eqnono(\displayt@g##1 )$$}}

\def\eqalignno#1{\displ@y \tabskip\centering
  \halign to\displaywidth{\hfil$\displaystyle{##}$\tabskip\z@skip
    &$\displaystyle{{}##}$\hfil\tabskip\centering
    &\llap{$\displayt@gpar##$}\tabskip\z@skip\crcr
    #1\crcr}}
\def\displayt@gpar(#1){(\displayt@g#1 )}
\def\displayt@g#1 {\rm\ifunc@lled{#1}\global\advance\tagnumber by1
        {\def\next{#1}\ifx\next\empty\else\expandafter
        \xdef\csname @qnnum#1\endcsname{\t@ghead\number\tagnumber}\fi}%
  \writenew@qn{#1}\t@ghead\number\tagnumber\else
        {\edef\next{\t@ghead\number\tagnumber}%
        \expandafter\ifx\csname @qnnum#1\endcsname\next\else
        \w@rnqwrite{  \noexpand\tag{#1} is   duplicate  }\fi}%
  \csname @qnnum#1\endcsname\fi}
\def\ifunc@lled#1{\expandafter\ifx\csname @qnnum#1\endcsname\relax}
\catcode`@=12
\def\toghead{{}}
 \hsize=5.75in
\vsize=8.75 in
\global\headline{\hfill}
\global\footline{\hfill\rm \folio\hfill}

\centerline{\sixteenbf QUANTUM GROUPS}
\medskip
\centerline{\bf M. Ruiz--Altaba}
\centerline{D\'ept. Physique Th\'eorique, %
Universit\'e de Gen\`eve, CH--1211 Gen\`eve 4}
\bigskip
\centerline{  UGVA--DPT--1993--10--838}
\centerline{  hep-th/9311069}
\bigskip

\centerline{ABSTRACT}
\medskip
These notes correspond rather accurately to
the translation   of the lectures given at the
Fifth Mexican School of Particles and Fields,
held in Guanajuato, Gto., in December~1992. They
constitute a brief and elementary introduction to
quantum symmetries from a physical point of view,
along the lines of the forthcoming book by
C.~G\'omez, G.~Sierra and myself.

\medskip
\medskip\line{  \hskip 1em\relax {}1. Introduction\hfil }
 \medskip\line{  \hskip 1em\relax {}2. Factorizable S--matrices\hfil }
 \medskip\line{  \hskip 1em\relax {}3. Bethe's diagonalization of %
 spin chain hamiltonians\hfil }
 \medskip\line{  \hskip 1em\relax {}4. Integrable vertex models:  %
 the six--vertex model\hfil }
 \medskip\line{  \hskip 1em\relax {}5. The Yang--Baxter algebra\hfil }
 \medskip\line{  \hskip 1em\relax {}9. Physical spectrum of the  %
 Heisenberg spin chain\hfil }
 \medskip\line{  \hskip 1em\relax {}6. Yang--Baxter algebras and  %
 braid groups\hfil }
 \medskip\line{  \hskip 1em\relax {}7.  Yang--Baxter algebras and %
 quantum groups\hfil }
 \medskip\line{  \hskip 1em\relax {}8. Affine quantum groups\hfil }
\medskip\line{  \hskip 1em\relax {}10. Hopf algebras \hfil}
\medskip\line{  \hskip 1em\relax {}11. The quantum group %
 $U_q({\cal G})$ \hfil}
\medskip\line{  \hskip 1em\relax {}12. Comments\hfil }
\medskip\line{  \hskip 1em\relax {}13. References\hfil }
\bigskip

\subchap{Introduction}
Since the advent of physics as a pleasure for the human  mind, those in
our trade have played  with  idealizations of reality which  preserve the
essence of the phenomenon and yet are simple enough to be modelled by the
mathematical tools available.  These ``toy models'', from the  point--like
friction-less particle in newtonian  mechanics to scalar quantum
electrodynamics, constitute most of the syllabus  of a physicist's
 education.  At the very worst, the theories and models we shall discuss
in these lectures can be  taken as paradigmatic toy models; indeed most of
two--dimensional field theory was invented  as a theoretical laboratory
for confinement, dimensional transmutation, instantons, and other such
 niceties of the real world. Yet something  deep remains   hidden in the
guts of  two--dimensional physics.  The historic  success of string theory
in unifying gauge symmetries  and general relativity has given the
two-dimensional world a new and fruitful life  of its own: from a stringy
point of view, enough remains to be learned about the fundamental
world--sheet that  we need not bother for a while about other dimensions.
Quite surprisingly, condensed matter  physicists have also come to
appreciate the interest of low--dimensional field theories,  motivated by
the planar character of the quantum hall effect and high--temperature
superconductivity or the technological  interest in thin plastic and
silicon chips, among other noteworthy phenomena.

These lectures are meant to illustrate some of the beautiful tricks that
can be applied to understanding two--dimensional physics.  Were we to
think of these lectures as a meal,  a menu would read more or less as
follows.  For appetizers, some  relativistic dynamics in one spatial
dimension.  The reader should then sit down to a light salad of
Bethe ansatz, followed by a hot soup of integrable vertex  models on the
plane. The main course consists of Yang--Baxter algebras, with a variety
of sauces of various mathematical origins. A few words about the possible
generalization of these treats to higher dimensions are left for desert.
All  the wines and liquors come from the quantum group vintage, distilled
at Kyoto and Kharkov from the well--known and now extinct  Leningrad
stock.  To avoid indigestion, the beautiful example of   two--dimensional
integrability provided by    conformal field theories  is not presented;
Professor Weyers's lectures in this same volume cover that.

 Quantum groups have been discovered relatively recently by physicists and
mathematicians concerned with integrable two--dimensional systems. An
integrable system has as many integrals of motion (constants) as
co-ordinates or, equivalently, momenta; accordingly, in an integrable
theory we know that the phase space is spanned by the so--called
action--angle variables, essentially a bunch of conserved quantities
(hamiltonians) and their conjugate variables (times).  Classical
two--dimensional statistical physics is equivalent to quantum field theory
in one (spatial) dimension. In these lectures we shall investigate
two-dimensional systems with an infinite number of degrees of freedom: we
shall be concerned with the thermodynamic limit of classical
two--dimensional statistical models.

 It is perhaps more intuitive to start with quantum field theories in one
space and one time dimension, that is quantum field theories on the real
line. Integrable field theories in such a small dimension is essentially
equivalent to the description of solitons. Indeed, a soliton is a
non--dispersive  classical solution to the classical equations of motion
which survives quantization and acquires a particle--like interpretation.
Feynman rules are, in general, rather useless in the description of
solitons. Collective phenomena of this sort are not perturbative at all,
and in low dimensions we might as well attack the problem directly to find
the $n$--point solitonic Green's functions.  In this endeavor,
integrability  comes in quite handy: we shall see shortly that the
existence of an infinity of conserved charges is equivalent to a
deceivingly simple  cubic equation in the $2\to2$ scattering amplitudes,
the celebrated Yang--Baxter equation.  In order to get a real theory, we
shall in addition impose unitarity and crossing symmetry.  Unitarity just
means that the probability is conserved, so that nothing comes of nothing
and  something comes of just as much.  Crossing symmetry is a more subtle
requirement, familiar from string theories, which can be viewed somehow as
a strong relativistic invariance, whereby particles moving forward or
backward in time can be traded off (suitably) by other particles moving
forward or backward in time.

\subchap{Factorizable S--matrices}
  We shall be interested in integrable  field theoretical  systems, \ie
systems with an infinite number of mutually commuting conserved charges.
One of these charges will be called the hamiltonian, an operator which
defines the time evolution of the system. To each conserved charge one can
associate a different time evolution: what we call the hamiltonian is a
matter of interpretation.

 Consider the scattering of relativistic massive particles in a
$(1+1)$--dimensional spacetime. There is only one spatial dimension   and
therefore the ordering of the particles is  well--defined, from  left to
right, say. In more spatial dimensions   we should not expect  that the
interesting features   depending strongly on the ordering of the
particles  remain valid.

  Introduce the rapidity $\theta$: $$  p^0 = m \cosh \theta
 \qquad,\qquad   p^1 = m \sinh \theta  \tag guagu$$ This parametrization
ensures the on--shell condition ${\vec p}^{\,2}=\left(   p^0  \right)^2 -
\left(   p^1  \right)^2  = m^2  $.

Alternatively, we could use the light-cone momenta $p$ and $\bar{p}$,
 $$  p= p^0 + p^1 = m\e^\theta \qquad,\qquad \bar{p}= p^0 - p^1 =
m\e^{-\theta} \tag$$  which transform under a  {Lorentz boost} $L_\alpha:
\theta\to \theta+\alpha$ as $$ p \to p \e^\alpha  \qquad,\qquad  \bar{p}
\to \bar{p} \e^{-\alpha}  \tag$$ Quite generally, an irreducible tensor
$Q_s$ of the  Lorentz group in $1+1$ dimensions is labelled by its spin
$s$ according to the rule $L_\alpha: Q_s \to \e^{s \alpha } Q_s  $, so
that $p$ is of spin~1 and its parity conjugate $\bar p$ is of spin~$-1$.

 If $Q_s$ is a local conserved quantity of integer spin $s>0$, then in a
scattering process involving  $n$ particles   $$\sum_{i\in\{{\rm in} \} }
p_i^s = \sum_{f\in\{{\rm out} \} } p_f^s  \tag $$ Similarly, if $Q_{-s}$
is conserved, then  $$\sum_{i\in\{{\rm in} \} } \bar{p}_i^s =
\sum_{f\in\{{\rm out} \} } \bar{p}_f^s  \tag $$ Setting   $s=1$ in \(+11)
and \(0), we recover  the usual   energy and momentum conservation laws of
a  relativistic   theory.

 The physical behavior of integrable systems is quite remarkable. For
 instance, if \(+11) and \(0) hold for an infinity of different spins $s$,
it follows immediately that the incoming and outgoing momenta must be the
same.  This means that no particle production or annihilation may ever
occur. Also,  particles with equal mass may reshuffle their momenta among
themselves in the scattering, but particles with different masses may not.
Equivalently, we may say that the momenta are conserved individually and
that particles of equal mass may interchange  additional  internal
quantum numbers. If all the incoming particles have different masses, then
the only effect of the scattering is a time delay (a  phase shift) in the
outgoing state with respect to the incoming one.

 All scattering processes can be understood and pictured as  a sequence of
two--particle scatterings. This property is called factorizability.

 By relativistic invariance, the scattering amplitude
 between two particles $A_i$ and $A_j$ may only depend on the scalar
$$p_i^\mu p_j^\nu \eta_{\mu\nu} = m_i m_j \cosh
\left(\theta_i-\theta_j\right)\tag $$ so that, in fact, it may depend only
on the rapidity difference $\theta_{ij}=\theta_i-\theta_j$. The general
form of the basic two--particle $S$--matrix   is
  $$ \ \ket{A_i(\theta_1), A_j (\theta_2) }  _{\rm in}  \longrightarrow
  \sum_{k,\ell} S_{ij}^{k\ell}(\theta_{12}) \ket{A_k(\theta_2), A_\ell
(\theta_1) }  _{\rm out} \tag mx110$$ In this notation,
$\ket{A_i(\theta_1), A_j (\theta_2) }  _{\rm in(out)}$ stands for the
initial (respectively, final) state of two incoming (respectively,
outgoing) particles of kinds $A_i$ and $A_j$ and rapidities $\theta_1$ and
$\theta_2$.

The second crucial feature of a factorizable  {$S$--matrix}
theory, from which such models get their name, is the property of
  {factorizability}:
the $N$--particle $S$--matrix can always be written as the product of
$N\choose2$ two--particle $S$--matrices.

We choose an initial state of $N$ particles with
 rapidities $\theta_1> \theta_2 > \cdots > \theta_N$ arranged in the
infinite past in the opposite order,  \ie $x_1<x_2 <  \cdots <x_N$. This
presumes simply that no scatterings may have occured before we study the
process, \ie that we have been looking long before any particles meet.
After the $N(N-1)/2$ pair collisions, the particles reach the infinite
future ordered along the spatial direction in increasing rapidity. Thus we
write
  $$\eqalignno{ S\ &\ket{A_{i_1}(\theta_1),\ldots, A_{i_N} (\theta_N) }
_{\rm in}  = \cr &\qquad = \;
 \sum_{j_1,\ldots,j_N} S_{i_1\cdots i_N}^{j_1\cdots j_N} \;
 \ket{A_{j_1}(\theta_N),\ldots, A_{j_N} (\theta_1) }  _{\rm out} &() \cr}
$$ Factorization  means that this process can be interpreted as a
set of independent and consecutive two--particle scattering processes.

 The spacetime picture of this multi-particle factorized scattering is
obtained by associating with each particle a line whose slope is the
particle's rapidity. The scattering process is thus represented by a
planar diagram with $N$ straight world--lines, such that no three ever
coincide at the same point. Any world--line will therefore intersect, in
general, all the other ones. The complete scattering amplitude associated
to any such diagram is given by the (matrix) product of two--particle
$S$--matrices.  For instance, in a four--particle scattering   we could
get  $$\eqalignno{ S_{i_1i_2 i_3 i_4}^{j_1 j_2 j_3 j_4} (\theta_1,
\theta_2, \theta_3, \theta_4) =
 \sum_{{k,\ell,m,n,\atop p,q,r,u}}
S_{i_1 i_2}^{k\ell} (\theta_{12})
S_{\ell i_3}^{mn} (\theta_{13})  \times & \cr  \times
S_{km}^{pq} (\theta_{23})
   S_{ni_4}^{rj_4} (\theta_{14})
S_{qr}^{uj_3} (\theta_{24})
S_{pu}^{j_1j_2} (\theta_{34}) & &(6)\cr}$$

 The kinematical data   (the rapidities of all the particles) does not fix
a diagram uniquely. In fact, for the same rapidites we have a whole family
of    diagrams, differing from each other by the  parallel shift of some
of the straight world--lines. The parallel  shift of any one line can (and
should) be  interpreted as a symmetry transformation. It corresponds to
the translation of the (asymptotic in-- and out--) $x$ co-ordinates of the
particle associated to the line.   Requiring the factorizability
condition   is equivalent to imposing that   the scattering amplitudes of
diagrams differing by such parallel shifts should be the same.

 For the simple case of three particles, the  condition that the
factorization be independent of  parallel shifts of the world--lines
amounts to  the following noteworthy factorization equation, which is the
necessary and sufficient  condition for any two diagrams differing by
parallel shifts to have equal associated amplitudes: $$ \eqalignno{
\sum_{p_1,p_2,p_3}  S_{i_1i_2}^{p_1p_2} (\theta_{12}) &S_{p_2i_3}^{p_3j_3}
(\theta_{13})  S_{p_1p_3}^{j_1j_2} (\theta_{23}) =
 \cr &=\sum_{p_1,p_2,p_3}
S_{i_2i_3}^{p_2p_3} (\theta_{23})
S_{i_1p_2}^{j_1p_1} (\theta_{13})
S_{p_1p_3}^{j_2j_3} (\theta_{12}) &(7)
\cr}  $$
 This is the famous
Yang--Baxter  equation.

To formalize this a bit more, consider a set of operators
$\left\{A_i(\theta) \right\}$  ($i=1,\ldots,n$) associated to each
particle $i$ with rapidity $\theta$, obeying the following commutation
relations: $$A_i (\theta_1) A_j (\theta_2) = \sum_{k,\ell} S_{ij}^{k\ell}
(\theta_{12}) A_k(\theta_2) A_\ell(\theta_1) \tag mx114$$ This equation
encodes   the two--particle scattering process \(mx110), where
``collision'' has been replaced by   ``commutation''. Furthermore, the
relation between \(mx110) and \(0) becomes evident if we interpret
$A_i(\theta)$ as an operator (Zamolodchikov operator) which creates the
particle $\ket{A_i(\theta)}$ when it acts on the Hilbert space vacuum
$\ket0$: $$A_i(\theta) \ket0 =\ket{A_i(\theta)} \tag mxff114$$

 The factorization equation \(+12) emerges in this context as a
``generalized
 Jacobi identity'' of the algebra \(+11), assumed associative.

 The following conditions are needed to guarantee the physical consistency
of the Zamolodchikov algebra \(mx114):

\list{{\it iii}}

\item{{\it i}}Normalization:
 $$\lim_{\theta\to0}  S_{ij}^{k\ell} (\theta) =  \delta_i^k \delta_j^\ell
\quad\iff\quad \lim_{\theta\to0}  S  (\theta) =\one \tag$$ This
condition   is obtained by setting $\theta_1=\theta_2$ in \(mx114). In
physical terms, it means that no scattering takes place if the relative
velocity of the two particles vanishes, \ie if the two world--lines are
parallel.

\item{{\it ii}}Unitarity:
$$\sum_{j_1,j_2}  S_{j_1 j_2}^{i_1i_2} (\theta) S_{k_1 k_2}^{j_1j_2}
(-\theta)  =  \delta_{k_1}^{i_1}   \delta_{k_2}^{i_2} \quad\iff\quad
   S  (\theta)  S  (-\theta) =\one \tag$$
This follows from applying \(mx114) twice.

\item{{\it iii}} Real analyticity:
 $$S^\dagger(\theta)=S(-\theta^*) \tag $$ which together with \(+11)
implies the physical unitarity condition $S^\dagger S=\one$.

\item{{\it iv}}{Crossing symmetry}:
 $$S_{ij}^{k\ell}(\theta) = S_{j {\bar\ell}   } ^{\bar{\imath} k}
(i\pi-\theta) \tag unimxmra$$ where  $\bar{\jmath} $ and $\bar k$ denote
the antiparticles of $j$ and $k$, respectively.

\endlist

 As an example, let us consider a theory with only one kind of particle
$A$ and its antiparticle $\bar{A}$. Due to {CPT invariance}, there exist
only three different amplitudes (we also assume conservation of  particle
number, \ie $\Zint_2$ invariance).  The scattering amplitude between
identical particles (or antiparticles) is denoted $S_I$, whereas $S_T$ and
$S_R$ denote the {transmission and reflection amplitudes}, respectively:
$$\eqalign{ & A(\theta_1) A(\theta_2) =  S_I (\theta_{12})  A(\theta_2)
A(\theta_1) \cr & A(\theta_1) \bar{A}(\theta_2) =  S_T (\theta_{12})
\bar{A}(\theta_2)  A(\theta_1)
 + S_R (\theta_{12})  A(\theta_2)  \bar{A}(\theta_1)\cr
 & \bar{A}(\theta_1) A(\theta_2) =  S_T (\theta_{12})  A(\theta_2)
\bar{A}(\theta_1)
 + S_R (\theta_{12})  \bar{A}(\theta_2)  A(\theta_1)\cr
 & \bar{A}(\theta_1) \bar{A}(\theta_2) =  S_I (\theta_{12})
\bar{A}(\theta_2) \bar{A}(\theta_1) \cr }\tag fal119$$
 It is not hard to check that the {factorization equations} for this
algebra read   as $$\eqalign{ & S_I S'_R S^{\prime\prime}_I = S_T S'_R
S^{\prime\prime}_T + S_R S'_I S^{\prime\prime}_R \cr   & S_I S'_T
S^{\prime\prime}_R = S_T S'_I S^{\prime\prime}_R +  S_R S'_R
S^{\prime\prime}_T \cr   & S_R S'_T S^{\prime\prime}_I = S_R S'_I
S^{\prime\prime}_T + S_T S'_R S^{\prime\prime}_R \cr  }\tag mx121$$
where   we have set $  S_a = S_a(\theta_{12})$, $  S'_a =
S_a(\theta_{13})$, $    S^{\prime\prime}_a = S_a(\theta_{23}) $ for
$a\in\{I,T,R\}$ to lighten the notation.

The  normalization conditions read   $$  S_I(0) =1
\quad,\quad  S_T(0)=0\quad,\quad  S_R(0)=1
 \tag$$ whereas  unitarity requires   $$\eqalign{ & S_T (\theta)
S_T(-\theta) + S_R (\theta) S_R(-\theta) =  1 \cr & S_T (\theta)
S_R(-\theta) + S_R (\theta) S_T(-\theta) =  0 \cr}\tag$$ and the
{crossing symmetry} implies $$S_I(\theta) = S_T(i\pi-\theta) \qquad ,
\qquad S_R(\theta) = S_R(i\pi-\theta) \tag$$

The equations \(mx121)  imply that the quantity  $$\Delta= { S_I(\theta)^2
+ S_T(\theta)^2 - S_R(\theta)^2  \over2 S_I(\theta) S_T(\theta) }  \tag
fal125$$ is   independent of the rapidity $\theta$.
An interesting factorized $S$--matrix is provided by the   sine--Gordon
theory, where the states $A$ and $\bar{A}$ of \(fal119) are identified
with the  {soliton} and antisoliton

\subchap{Bethe's diagonalization of spin chain hamiltonians}
Consider now a periodic one--dimensional regular lattice (a periodic chain)
with $L$ sites. At each site, the spin variable may be either up or down,
so that the Hilbert space of the spin chain is simply ${\cal H}^{(L)}  =
\bigotimes^L V^{\12} $ where $V^\12$ is the spin--$\12$ irreducible
representation of $SU(2)$ with basis $\{ \ket\Uparrow, \ket\Downarrow\}$.
By simple combinatorics, the dimension of the Hilbert space is dim~$ {\cal
H}^{(L)}  = 2^L$. On ${\cal H}^{(L)}$, we consider a very general
hamiltonian $H$, subject to three constraints.

 First, we assume that the  interaction is of short range, for example only
among nearest neighbors.   Next, we impose that the hamiltonian $H$ be
translationally invariant. Letting  $  \e^{iP} $ denote the operator which
shifts the states of the  chain by one lattice unit to the right, then
this requirement reads as $\left[ \e^{iP} , H \right] =0 $.      From
periodicity of the closed chain, we must have $ \e^{iPL}=\one $ Finally,
we demand that the hamiltonian  preserve the third component of the spin:
$$\left[ H, S^z_{\rm total} \right] =  \left[ H, \sum_{i=1}^L S^z_i
\right] = 0 \tag mx131$$ This requirement allows us to divide the Hilbert
space of states into different sectors, each labelled by the third
component of the spin or, equivalently, by the total number of spins down.
We shall denote by ${\cal H}^{(L)}_M$ the subspace of ${\cal H}^{(L)}$
with $M$ spins down.  Obviously, dim~$ {\cal H}^{(L)}_M  = {L \choose M}$,
so that dim~$ {\cal H}^{(L)}  = \sum_{M=0}^L {\rm dim}\; {\cal
H}^{(L)}_M$.

 We wish to study  the eigenstates and spectrum of $H$. The zero-th sector
${\cal H}_0^{(L)}$ contains only one state, the ``{Bethe reference
state}'' with all spins up. The most natural ansatz for the eigenvectors
of $H$ in the other sectors is some superposition of ``{spin wave}s'' with
different velocities. For the  first sector, \ie the subspace of states
with all spins up except   one down, the ansatz for the eigenvector is
thus of the form  $\ket{\Psi_1} = \sum_{x=1}^L f(x) \ket{x}  $ where $\ket
x$ represents the state with all spins up but for the one at lattice site
$x$ ($1\le x\le L$).  The unknown wavefunction $f(x)$ determines the
probability that the single spin down is precisely at site~$x$.

{}From   the complete  translational invariance due to periodic boundary
conditions, it is reasonable to  assume that $f(x)$ is just the
wavefunction for a plane wave   $$f(x)= \e^{ikx} \tag mx133$$ with some
particular momentum $k$ to be fixed by the boundary condition $ f(x+L)
=f(x)  $.   Thus $k=2\pi I/L$, with $I=0,1,\ldots, L-1$. Hence the
eigenvectors of $H$ with one spin down   span indeed  a basis of the
Hilbert space ${\cal H}^{(L)}_1$, by dimensionality counting.

The wavefunction solving the eigenvalue problem for the sector with two
spins down,  $H\ket{\Psi_2} = E_2 \ket{\Psi_2} $,   is  of the form
 $\ket{\Psi_2}= \sum_{x_1, x_2}  f(x_1,x_2) \ket{x_1,x_2} $, where
$\ket{x_1,x_2} $ stands for the state with ll spins up except two spins
down at positions $x_1$ and $x_2$.

 The   {periodicity condition}  reads now  $f(x_1, x_2) = f(x_2, x_1+L)
$.   The most naive ansatz for $f(x_1, x_2)$ generalizing the plane wave
is  $f(x_1, x_2) = A_{12} \;  {\rm exp}\,{i(k_1x_1 + k_2 x_2)}  $. This
ansatz is inappropriate, however, because it violates the periodicity
condition. Physically, we have forgotten to include the scattering of the
two ``spin waves'' with ``quasi-momenta'' $k_1$ and $k_2$. The solution to
this problem was found by Bethe, who wrote the useful ansatz     $$f(x_1,
x_2)=A_{12} \e^{i(k_1x_1 + k_2 x_2)} + A_{21} \e^{i(k_1x_2 + k_2 x_1)}
\tag  fal139$$ which does satisfy the periodicity condition   provided
$$ A_{12} = A_{21} \e ^{ik_1 L}  \quad,\qquad A_{21} = A_{12} \e ^{ik_2
L}   \tag$$ Note that these two conditions imply, in particular, that
$\exp i(k_1+k_2)L=1$, which reflects the invariance of the wavefunction
under a full turn around the chain, \ie under the shift of $L$ units of
lattice space: $$f(x_1+L, x_2+L)= f(x_1, x_2) \iff \e^{ i(k_1+k_2)L} =1
\tag$$ This equation must hold if the wavefunction is to be single valued.

 The ansatz \(fal139) already assumes that the $S$--matrix for two spin
waves is purely elastic. In fact, the only dynamics   allowed is the
permutation of the quasi-momenta.

 To capture the physical meaning behind equations \(+11), let us introduce
the ``scattering amplitudes for  {spin wave}s''
$$  \hat{S}_{12} = {A_{21}\over A_{12} }   \quad,\qquad
 \hat{S}_{21}= {A_{12}\over A_{21} }   \tag mx1355$$ in terms of which
\(+12) read as  $$  \e^{ i k_1 L}  \hat{S}_{12}(k_1, k_2) =1 \quad,\qquad
\e^{ i k_2 L}  \hat{S}_{21}(k_2, k_1) =1  \tag mx1356$$  These equations
tell us that the total   {phase shift} undergone by a spin wave after
travelling all the way around the closed chain is one. This phase shift
receives two contributions; one is purely kinematic ($\e^{ik_1L}$  or
$\e^{ik_2L}$) and depends only on the   {quasi-momentum} of the spin
waves, while the other reflects the phase shift produced by the
interchange of the two spin waves.

 Summarizing the previous discussion, we have found that the  Bethe ansatz
for the eigenvector of the hamiltonian $H$ in the sector $M=2$ is
 $$f(x_1, x_2) = A_{12} \left( \e^{i(k_1x_1 + k_2 x_2)} +
\hat{S}_{12}(k_1, k_2) \e^{i(k_1x_2 + k_2 x_1)}  \right) \tag micifus$$

  The generic form of a state $\ket{\Psi_M}\in {\cal H}^{(L)}_M$ in
the sector with
$M>2$ spins down is
 $$\ket{\Psi_M} = \sum_{1\le x_1<x_2<\cdots<x_M\le L}  f(x_1, \ldots,x_M)
\ket{x_1, \ldots,x_M} \tag genricus$$ The Bethe ansatz is now
 $$f(x_1,  \ldots,x_M) = \sum_{p\in {\cal S}_M} A_p \e^{i(k_{p(1)}x_1 +
\cdots + k_{p(M)} x_M)} \tag mx13510$$ where the sum runs over the $M!\;$
permutations $p$ of the labels of the
 quasi-momenta $k_i$.    The periodicity condition is now $$f(x_1,x_2,
\ldots,x_M) =f(x_2,  \ldots,x_M, x_1+L)  \tag mx13511$$

When $M=3$,   we get the following six equations:
$$    \e^{ik_1L} = {A_{123} \over A_{231} }  = {A_{132} \over A_{321} }
\quad , \quad  \e^{ik_2L} = {A_{231} \over A_{312} }  = {A_{213} \over
A_{132} }  \quad , \quad  \e^{ik_3L} = {A_{312} \over A_{123} }  =
{A_{321} \over A_{213} }    \tag$$ Thus, in addition to the relations
among the quasi-momenta $k_i$ and the amplitudes $A_p$, there exist
additional constraints among the amplitudes of three quasi-particles,
which were absent in the simpler case with $M=2$.  These relations   tell
us that the interchange of two particles is independent of the position of
the third particle. Locality of the interactions is thus equivalent to the
 {factorization} property of the $S$--matrix, according to which the
scattering amplitude of $M$ quasi-particles factorizes into a product of
${M \choose2}$ two--point $S$--matrices.

 The Yang--Baxter content of the Bethe ansatz for $M=3$ is illustrated
with the following equalities: $$A_{321}=\cases{ \hat{S}_{12}\; A_{312} =
\hat{S}_{12}\hat{S}_{13}\;  A_{132} =
\hat{S}_{12}\hat{S}_{13}\hat{S}_{23}\;  A_{123} &\cr \hat{S}_{23}\;
A_{231} = \hat{S}_{23}\hat{S}_{13}\;  A_{213} =
\hat{S}_{23}\hat{S}_{13}\hat{S}_{12}\;  A_{123} &\cr} \tag mx13514$$

We thus   arrive  to the all--important
``{Bethe ansatz equations}''
 $$\e ^{ i k_i L} = \prod _{{j=1\atop j\not= i}}^M \hat{S} _{ji} (k_j,
k_i)  \qquad{\rm for}\;\; i=1,\ldots,M \tag mx13515$$ written in general
for a sector with arbitrary $M$.  The actual solution to these equations
far transcends the framework of these lectures. Suffice it to say that
 a variety of methods have been devised to attack them.

 The spin wave scattering amplitude  $\hat{S}_{12}$ depends of course on
the detailed form of the hamiltonian, and it can be computed by solving
the $M=2$ eigenvalue equation, which reads more explicitly as  $$E_2 \;
f(x_1, x_2) = \sum_{1\le y_1 <y_2\le L}  \bra{x_1, x_2} H \ket{y_1, y_2}
f(y_1, y_2) \tag mx1358$$ Using \(micifus) in \(0), we would find
$\hat{S}_{12}$ as a function of $k_1$, $k_2$ and the matrix elements of
$H$.

 Unfortunately, there does not exist a simple criterion to decide when a
{spin chain} hamiltonian is  integrable, \ie when it allows the Bethe
construction. As we have shown, however,   the Bethe ansatz will work
whenever the spin wave $S$--matrix satisfies the integrability
condition    and  factorization.    Let us stress   that the
diagonalization of a hamiltonian with the help of the Bethe ansatz does
not even work for any translationally invariant and   short range
hamiltonian preserving the total spin.  Only a very special class of such
hamiltonians can be diagonalized via the Bethe procedure, namely those
which describe integrable models.  An important spin chain model, to which
the Bethe ansatz technique is   applicable, is the  {$XXZ$ model}
$$H_{XXZ} = J \sum_{i=1}^{L} \left(  \sigma_i^x \sigma_{i+1}^x +
\sigma_i^y \sigma_{i+1}^y + \Delta \sigma_i^z \sigma_{i+1}^z  \right)\tag
fal153$$

\subchap{Integrable vertex models: the  six--vertex model}
 Let us now turn to  classical  statistical systems  in two spatial
dimensions  (in equilibrium, so no time dimension)  on a  lattice. Whereas
in the previous section the main problem consisted in diagonalizing a
one--dimensional hamiltonian, in this section we address the computation
of the  partition function of the lattice system.

A {vertex model} is a statistical model defined on a  lattice ${\cal L} $,
taken  regular  and rectangular for simplicity.  We shall thus consider an
$L\times L'$   lattice with $L$ vertical lines (columns) and $L'$ horizontal
lines (rows).

 A physical state on this lattice is defined by the assignment to each
lattice edge of  a state variable, characterized by some labels; allowing
for two possible states on each link suffices for our purposes. These two
possiblities may be interpreted as spins up or down.  Alternatively,   if
we imagine the lattice links as electric wires with a current of constant
intensity running through them, then the two states are associated with
the direction of the current.

 The dynamics of the model is   characterized by the interactions among
the lattice variables, which take place at the vertices, whence the name
vertex model.  The energy $\varepsilon_V$ associated with a vertex $V$
depends only on the four states on the edges meeting at that vertex
(locality). This is also true for the    Boltzmann weights $W_V=
\exp\left(  -\varepsilon_V /k_BT \right)$ which measure the probability of
each local configuration. Quite generally, it is convenient to represent
{Boltzmann weight}s as $ W\pmatrix{ \beta& \nu\cr \mu& \alpha\cr}$, where
$\mu$ and $\nu$ are the horizontal edge state labels, and $\alpha$ and
$\beta$ the vertical ones: $$W\pmatrix{ \beta& \nu\cr \mu&
\alpha\cr}\qquad = \qquad \mu {\qquad {\displaystyle \beta\atop\Bigg|}
\qquad \over {\Bigg| \atop \displaystyle \alpha} } \nu \tag$$

  If we impose that the interaction conserve  the  total spin   or the
local current, then all but six Boltzmann weights must vanish.   In
addition to the spin, particle number or current conservation, we may also
impose the  $\Zint_2$ {reversal symmetry} under $\ket\Uparrow
\leftrightarrow \ket\Downarrow$. Under this condition, the   independent
Boltzmann weights are reduced to three, which we shall call just $a$, $b$
and $c$. These  weights define the symmetric or zero--field {six--vertex
model}, which we shall call the six--vertex model for short.

The six--vertex model is characterized by   link variables $\in
{\Zint_2}=\{0,1\}$ with Boltzmann weights subject to current conservation
 $$ W\pmatrix{ \beta& \nu\cr \mu& \alpha\cr} = 0  \qquad {\rm unless}
\qquad \mu+\alpha= \nu+\beta \tag mx151a$$ and reflection symmetry
($\bar{x}=1-x$) $$ W\pmatrix{ \beta& \nu\cr \mu& \alpha\cr} =W\pmatrix{
\bar\beta& \bar \nu\cr \bar \mu& \bar\alpha\cr} \tag$$
A   compact way to write the weights   is
 $$ W\pmatrix{ \beta& \nu\cr \mu& \alpha\cr}\;= \; b\;
\delta_{\mu\nu}\,\,  \delta _{\alpha\beta}\; +\; c \; \delta_{\mu\beta}
\,\, \delta_{\nu\alpha} \;+\; (a-b-c) \;  \delta_{\mu\alpha} \,\, \delta
_{\nu\beta} \tag mx152$$

The  partition function is $$Z_{L\times L'} (a,b,c)= \sum_{  {\cal C} } \e
^{- E({\cal C}) /k_BT} = \sum_{  {\cal C} } \prod_V W_V \tag mx153$$ where
the sum runs over all possible configurations $\cal C$, of which there are
$2^{LL'}$. In the {thermodynamic limit}, when $L$ and $L'$ tend to
infinity, the computation of the sum \(0) becomes a rather formidable and
apparently insurmountable problem.   Lieb's breakthrough to compute  the
partition function \(0) of the six--vertex model  relies basically on
rephrasing the problem as  the diagonalization of the anisotropic
spin--$\12$ chain, which had been solved already by the  Bethe ansatz.
First, let us perform the sum over the horizontal variables, which
involves only the Boltzmann weights on the same row of the lattice, and
then carry out the sum over the vertical variables.  The double  sum \(0)
can thus be rearranged as follows: $$  Z_{L\times L'} (a,b,c) =
\sum_{{{\rm vertical }\atop{\rm states}}}  \prod_{\rm rows}  \left(
\sum_{{{\rm horizontal}\atop{\rm states}}}  \prod_{V\in {\rm row}} W_V
\right) \tag $$ The quantity in parenthesis depends  on the two sets of
vertical states above  and below the row of horizontal variables: it is
the (row to row) transfer matrix of the model. For conceptual clarity, it
is convenient to introduce the ``fixed time states'' as the set of
vertical link variables on the same row: $$\ket\alpha = \matrix{ \alpha_1
\cr \Big|\cr\cr} \quad  \matrix{ \alpha_2 \cr \Big|\cr\cr} \quad  \matrix{
\alpha_3 \cr \Big|\cr\cr} \quad \cdots \quad  \matrix{ \alpha_{L-1} \cr
\Big|\cr\cr} \quad  \matrix{ \alpha_L \cr \Big|\cr\cr} \tag$$ The transfer
matrix element $\vev{\beta |t|\alpha}$ can then be understood as the
transition  probability for the state $\ket\alpha$ to project on the state
$\ket\beta$ after a unit of time. We are thinking now of the horizontal
direction as space, and the vertical one as time: $$\eqalignno{ \bra\beta
t(a,b,c) \ket\alpha = \sum_{\mu_i} &
 W\pmatrix{ \beta_1& \mu_2\cr \mu_1& \alpha_1\cr}
 W\pmatrix{ \beta_2& \mu_3\cr \mu_2& \alpha_2\cr} \cdots  \cr
&\cdots  W\pmatrix{ \beta_{L-1}& \mu_L\cr \mu_{L-1}& \alpha_{L-1}\cr}
 W\pmatrix{ \beta_L& \mu_1\cr \mu_L& \alpha_L \cr}& (mx156)\cr}   $$
 We agree with the Chinese who think that a picture is better than a
formula: $$\vev{\beta |t|\alpha}\quad =\quad \sum_{\mu_i}  \quad{ \quad
{\displaystyle \beta_1 \atop\Bigg|} \quad
 {\displaystyle\beta_2 \atop\Bigg|}  \qquad  \cdots \qquad
 {\displaystyle\beta_{L-1} \atop\Bigg| } \quad
 {\displaystyle\beta_L \atop\Bigg| } \quad \over
  {\Bigg|\atop\displaystyle \alpha_1 } \quad
{\Bigg|\atop\displaystyle\alpha_2 }
\qquad \cdots \qquad  {\Bigg|\atop\displaystyle\alpha_{L-1}} \quad
{\Bigg|\atop\displaystyle\alpha_L} } \hskip-6.2cm
{ \quad \atop \mu_1 \quad  \mu_2 \quad  \mu_3 \qquad\quad  \mu_{L-1}
\qquad \mu_L \quad     \mu_1  }  \tag$$
 The {transfer matrix} $t(a,b,c)$ plays the role of a discrete evolution
operator acting on the Hilbert space ${\cal H}^{(L)}$ spanned by the row
states $\ket\alpha$  (dim~${\cal H}^{(L)} =2^L$), isomorphic to the one
considered above in the   diagonalization of the spin--$\12$ hamiltonian.
The full  partition function reads thus $Z_{L\times L'} (a,b,c) = {\rm
tr}_{{\cal H}^{(L)}}  \left( t(a,b,c) \right)^{L'}$.  The  trace on
${\cal H}^{(L)}$ implements   periodic boundary conditions in the ``time''
direction.  This expression is just the hamiltonian formulation of the
partition function  \(mx153). Thus  evaluating  the partition function is
in fact equivalent to finding the eigenvalues of the transfer matrix. We
are led, therefore, to essentially the same problem considered in the
previous section, namely the diagonalization of an operator on ${\cal
H}^{(L)}$.

First of all, the local conservation law \(mx151a) translates into
$\bra\beta t \ket\alpha =0$ unless the total spin is equal for both
$\ket\alpha$ and $\ket\beta$, $\sum_{i=1}^L \alpha_i =  \sum_{i=1}^L
\beta_i  $. More technically, the {number operator} $M= \sum_{i=1}^L
\alpha_i$ commutes with the {transfer matrix}:  $$ \left[ t(a,b,c) , M
\right] =0 \tag$$ This is the analog of equation \(mx131) and the relation
between the total spin $S^z$ and $M$ is simply   $S^z = {L\over 2} -M $.
Once again, the Hilbert space ${\cal H}^{(L)}$ can be broken down into
sectors ${\cal H}_M^{(L)}$ labelled by $M\in\{0,1,\ldots,L\}$. In each of
these sectors, the transfer matrix can be diagonalized independently,
$t(a,b,c) \ket{\Psi_M} = \Lambda_M(a,b,c) \ket{\Psi_M} $.  The states
$\ket{x_1,\ldots,x_M}$ with 1's at the positions $x_1,\ldots, x_M$ and 0's
elsewhere form a basis of ${\cal H}_M^{(L)}$  Expanding $\ket{\Psi_M}$ in
this basis,   $$\ket{\Psi_M} = \sum_{ 1\le x_1 < x_2 < \cdots < x_M \le L}
f(x_1, \ldots, x_M)  \ket{x_1, \ldots, x_M} \tag$$ we find the equation
for the eigenfunctions $f(x_1, \ldots, x_M)$:  $$ \eqalignno{  \sum_{ 1\le
y_1 < y_2 < \cdots < y_M \le L} & \bra{x_1, \ldots, x_M} t(a,b,c)
\ket{y_1, \ldots, y_M} f(y_1, \ldots, y_M) \cr & = \Lambda_M(a,b,c)
f(x_1,\ldots, x_M) &()\cr}  $$     The transfer matrix connects states
with the same number $M$ of down spins, whose locations may change.  The
eigenvalue problem \(0) can be solved with the help of the Bethe ansatz
technique.

 The sector $M=0$ contains only one state $\ket\Omega=\ket{00\ldots0}$
which is  the {Bethe reference state}. This state plays the role of a
vacuum in the construction of the other states, but it   need not coincide
with the  ground state of the model: the physical vacuum  minimizes the
{free energy} and may have nothing to do with $\ket\Omega$.  From \(mx156)
and \(0), we obtain  $$  \Lambda_0  = \bra\Omega t \ket\Omega =
\sum_{\mu=0,1} \left[ W \pmatrix{ 0& \mu\cr \mu& 0\cr} \right]^L  = a^L
+  b^L  \tag$$

 In the sector $M=1$, we choose $f(x)=\e^{ikx}$ and some elementary
algebra yields (with the assumption of   {periodic boundary conditions})
$$\Lambda_1(k) = a^L P(k) + b^L Q(k) \tag$$
with $$  P(k) = {ab+ (c^2 -b^2 ) \e^{-ik} \over a^2 -ab \e^{-ik} }
\quad, \quad Q(k) = {a^2-c^2 -ab \e^{-ik} \over ab -b^2 \e^{-ik} }
  \tag mr182$$

 The sector with $M=2$ excitations contains more structure. The  {Bethe
ansatz} reads in this case $$ f(x_1, x_2 ) = A_{12} \e^{i ( k_1 x_1 + k_2
x_2 )} + A_{21}  \e^{i ( k_2 x_1 + k_1 x_2 )} \tag$$ subject to the
periodic boundary conditions,  which yields the equations \(mx1355) and
\(mx1356). The eigenvalue of \(0) is given by $$\Lambda_2 = a^L P_1 P_2 +
b^L Q_1 Q_2 \tag mr185$$ where $P_i= P(k_i)$ and $Q_i = Q(k_i)$. The ratio
of the amplitudes $A_{12}$ and $ A_{21}$, which again may be interpreted
as a  {spin wave} scattering matrix, is thus  $$ \hat{S}_{12} = {A_{21}
\over A_{12}} = -\ { 1-2 {a^2 +b^2 -c^2 \over 2ab} \e^{ik_2} +
\e^{i(k_1+k_2)}  \over  1-2 {a^2 +b^2 -c^2 \over 2ab} \e^{ik_1} +
\e^{i(k_1+k_2)} } \tag mxmx$$ For later convenience, we define  the
{anisotropy parameter} $\Delta$ as  $$\Delta = {a^2 +b^2 -c^2 \over
2ab}\tag mx169$$ Notice the strong similarity between \(0) and \(fal125).

The generalization of the above results to  sectors with more than two
excitations proceeds through the  factorization properties of the higher
order Bethe amplitudes $A_{1\cdots M} $ [see equations \(mx13514)]. The
general formula for the eigenvalue $\Lambda_M$ of a vector of the form
\(mx13510) is $$\Lambda_M = a^L \prod_{i=1}^M P(k_i) + b^L \prod_{i=1}^M
Q(k_i) \tag mx170$$ and the  quasi-momenta $k_i$ ($i=1,\ldots,M$) must
satisfy the Bethe equations  \(mx13515)  which follow from the periodicity
\(mx13511) of the wave functions and the factorization properties of the
Bethe amplitudes. In this case, they read explicitly as $$\e^{ik_iL} =
(-1)^{M-1} \prod_{{j=1 \atop j\not= i }}^M { 1-2 \Delta \e^{ik_i} +
\e^{i(k_i+k_j)}  \over  1-2 \Delta \e^{ik_j} + \e^{i(k_i+k_j)} } \tag
mx171$$

The final step in the computation of the eigenvalues of the transfer
matrix  and, ultimately, of the  partition function, hinges upon the
solution of the Bethe equations \(mx171). It is very important  that the
Bethe equations associated with the  {six--vertex model} depend on the
Boltzmann weights $a$, $b$ and $c$ only through the combination yielding
the anisotropy $\Delta$ in  \(+12). This is the key for understanding the
integrability of the six--vertex model. The first immediate consequence
from this observation is    that two different transfer matrices $t(a,b,c)
$ and $t(a',b',c')$  sharing the same value for $\Delta$ have the same
eigenvectors and thus they commute:  $$ \left[ \;  t(a,b,c) \;   , \;
t(a',b',c')  \;  \right] \;   = \;  0 \quad \iff\quad \Delta(a,b,c)=
\Delta(a',b',c') \tag mx172$$ Therefore, given a value of $\Delta$,
through the Bethe procedure we diagonalize not just a  transfer matrix but
a whole continuous family of mutually commuting transfer matrices. Since
each   transfer matrix defines a different time evolution, to each
transfer matrix with the same parameter $\Delta$ is associated   a
conserved quantity. So the six--vertex model has a large number of
conserved quantities, in fact an infinity of them in the thermodynamic
limit.

 Let us sketch now the relation between the one--dimensional hamiltonian
 $H_{XXZ}(\Delta)=H_\Delta$ of the anisotropic  Heisenberg chain
\(fal153)  and the two--dimensional  six--vertex model. From the
identification of the Bethe ansatz eigenvectors under \(+13), we see that
$$ \left[ \;   H_{\Delta(a,b,c)} \;    , \;  t(a,b,c)  \;  \right] \;   =
\;  0 \tag$$   Comparison of this expression with \(+11) leads us to
suspect that the  hamiltonian $H_\Delta$ must already be contained somehow
in the  transfer matrix $t(a,b,c)$, \ie it should be one of the conserved
quantities in the system.  The same argument applies also to the
translation operator $\e^{iP}$ which commutes both with the hamiltonian
and with the transfer matrix.

To make these suggestive connections explicit, let us start with the
momentum operator   $\e^{-iP}$. Suppose we make the following choice of
Boltzmann weights: $$ a=c=c_0 \qquad,\qquad b=0 \tag rafpra110$$ which is
consistent with any value of $\Delta$. Then from \(mx152) we get
$$W\pmatrix{ \beta&\nu\cr \mu&\alpha\cr} \Bigg|_{{a=c=c_0 \atop b=0}} =
c_0 \; \delta_{\mu\beta}\; \delta_{\nu\alpha} \tag$$ which can be imagined
  as an operator which multiplies by $c_0$   the incoming
state $\{\mu,\alpha\}$ but otherwise leaves it untouched: the horizontal
state on the left becomes the vertical state on top, and the vertical
state below becomes the horizontal state to the right.  Thus the
{transfer matrix}  from  these weights behaves as the {shift operator}
$\e^{-iP}$: $$\eqalign{ t_0\ket\alpha\ & = t(c_0,0,c_0) \ket{\alpha_1,
\alpha_2, \ldots, \alpha_L} \cr &= c_0^L \quad\ket{ \alpha_L , \alpha_1,
\ldots , \alpha_{L-1}} \cr}\tag$$  and the  {momentum operator} $P$ is
identified with $$P= i\log \left( t_0 \over c_0^L \right)  \tag mx176$$
This identification is easily checked on one--particle states. From
\(+11), we see that $t_0\ket x= \ket {x+1}$, and thus in the Fourier
transformed states $\ket k$ we find  $$\eqalign{ t_0 \ket k\;& =  t_0
\sum_{x=1}^L \e^{ikx} \ket x = \sum_{x=1}^L \e^{ikx} \ket{ x+1}\cr & =
\sum_{x=1}^L \e^{ik(x+1)} \e^{-ikx} \ket{ x+1} = \e^{-ik} \ket k \cr}
\tag$$

 Similarly, the hamiltonian $H_\Delta$ can be obtained by expanding the
transfer matrix in the vicinity of the parameter point \(rafpra110),
keeping the value of $\Delta $ constant. Note that this amounts to
expanding the    transfer matrix about $t_0$, \ie about a matrix
proportional to the shift operator. So we fix $$\Delta = {\delta a -
\delta c \over \delta b } \tag$$ and obtain  $$t_0^{-1}\;  \delta t =
{\delta b \over 2c_0}\quad \sum_{i=1}^L \left\{ {\delta a + \delta c \over
\delta b}\;  \one + \sigma_i^x \sigma_{i+1}^x + \sigma_i^y \sigma_{i+1}^y
+  \Delta \sigma_i^z \sigma_{i+1}^z \right\} \tag fal191$$ Thus the
hamiltonian $H_\Delta$  [see \(fal153)]  appears in the expansion of the
logarithm of the transfer matrix about the shift operator,  $$  H_\Delta=
i {\partial\over \partial u} \log { t(u) \over c_0^L } \Big|_{u=0}
\tag$$  We urge the reader to carry out the above simple and most
instructive calculation explicitly.

 Expanding $\log t(u)$ in powers of $u$ we  get a whole set of  local
conserved quantities    involving in general interactions over a finite
range, not just among nearest neighbors.  In this sense, the  transfer
matrix is the generating functional for a large class of commuting
conserved quantities: this follows from the integrability  equation
\(mx172). Recall that we arrived at that equation after a long analysis
involving the Bethe ansatz. Instead, we could have taken equation \(mx172)
as the starting point to define a  {vertex model}, and asked ourselves
under what conditions the Boltzmann weights of such  vertex model lead to
integrability. The answer to this question is that, in order that the
vertex model be integrable, the Boltzmann weights must satisfy the justly
celebrated Yang--Baxter equation:  $$\eqalignno{ \sum_{\mu',\nu',\gamma}
&W\pmatrix{ \gamma & \mu' \cr \mu & \alpha \cr}  W'\pmatrix{ \beta& \nu'
\cr \nu& \gamma \cr}   W^{\prime\prime}\pmatrix{ \nu^{\prime\prime}&
\mu^{\prime\prime} \cr \nu'& \mu' \cr} =\cr &= \sum_{\mu',\nu',\gamma}
W^{\prime\prime}\pmatrix{ \nu' & \mu' \cr \nu & \mu \cr}  W'\pmatrix{
\gamma& \mu^{\prime\prime} \cr \mu'& \alpha \cr}   W \pmatrix{ \beta&
\nu^{\prime\prime} \cr \nu'& \gamma  \cr} &(ybchap1)\cr}$$  where $W$,
$W'$ and $W^{\prime\prime}$ are three different sets of Boltzmann
weights.

 Equation \(0) is the {Yang--Baxter equation for vertex  models}.  Note
that in the two problems worked out so far, the diagonalization of  {spin
chain hamiltonian}s and the diagonalization of the transfer matrix for
vertex models, integrability is encoded in the same mathematical structure,
namely the factorization of the spin wave $S$--matrix \(mx13515) and the
vertex Yang--Baxter equation \(0). There exist several other formulations
of the Yang--Baxter equation, which is always   a cubic equality;  it
first appeared under the name of  {star--triangle relation} in Onsager's
solution to the two--dimensional  {Ising model}.

A side comment: in statistical mechanics,  the local energies must be real
and thus the Boltzmann weights must be real and positive.   It is
nevertheless useful to allow the Boltzmann weights to be complex in
general. This freedom is useful from the technical viewpoint, but it is
also physically meaningful. In particular, the region of parameter space
in which the one--dimensional spin chain hamiltonian is hermitian need not
coincide with that in which the two--dimensional Boltzmann weights are
real and positive. Thus the physical spin chain hamiltonian is indeed an
analytic extension of the hamiltonian derived from   realistic  Boltzmann
weights.

\subchap{The Yang--Baxter algebra}
 Let us analyze in more detail the structure of the Yang--Baxter equation
\(ybchap1).  Our goal is to capture in a general algebraic framework  the
integrability properties of the models studied above.  The  transfer
matrix is an   endomorphism $$t(a,b,c) : V_1\otimes \cdots \otimes V_L \to
V_1\otimes \cdots \otimes V_L \tag $$ where $V_i$ stands for the
spin--$\12$ representation space at the $i$--th position of the lattice.
This operator is built up by multiplying {Boltzmann weight}s on the same
row and summing over the horizontal states connecting them, while keeping
the vertical states above and below them fixed [see eq. \(mx156)]. To make
this distinction even clearer, we shall refer to the space of horizontal
states as auxiliary, and  denote it by $V_a$.   The space of vertical
states on which the {transfer matrix} acts we shall call quantum and
denote it as ${\cal H}^{(L)}=V_1\otimes \cdots \otimes V_L$, as in \(0).
For the {six--vertex model}, $V_a$ is also a spin--$\12$ representation
space, like $V_i$ ($i=1,\ldots,L$).

According to these  definitions, it is natural to interpret the Boltzmann
weights associated to the $i$--th vertex as an operator $\cR _{ai}$:
 $$\cR _{ai}: V_a \otimes V_i \to V_a \otimes V_i \tag $$ where the
subindices in $\cR$ label the vector spaces it acts upon. The operator
$\cR_{ai}$ is  defined by its matrix elements  $$\eqalign { \mu_i {\quad
{\displaystyle \beta_i \atop \Big| } \quad \over {\Big|
\atop\displaystyle  \alpha_i } } \mu_{i+1} \  &= W \pmatrix{ \beta_i &
\mu_{i+1} \cr \mu_i & \alpha_i\cr} \equiv  \cR _{\mu_i \alpha_i}
^{\mu_{i+1}  \beta_i} \cr &= \,_a\bra{\mu_{i+1}} \otimes \,_i
\bra{\beta_i}\; \cR _{ai} \; {\ket{\mu_i}}_a \otimes  {\ket{\alpha_i}}_i
\cr}\tag$$  Note that if $\cR$ appears with two subindices, they label the
spaces $\cR$ acts upon, whereas if $\cR$ appears with two subindices and
two superindices, they label the basis vectors of the spaces $\cR$ is
acting between. Using these notations, the transfer matrix \(mx156) can be
written as  $$\bra\beta t \ket\alpha = \sum_{\mu'{\rm s}} \cR _{\mu_L
\alpha_L}^{\mu_1 \beta_L}   \cR _{\mu_{L-1} \alpha_{L-1}}^{\mu_L
\beta_{L-1}}  \cdots  \cR _{\mu_2 \alpha_2}^{\mu_3 \beta_2}
 \cR _{\mu_1 \alpha_1}^{\mu_2 \beta_1} \tag mx24$$ We have reversed the
order of multiplication of the Boltzmann weights to agree with the
 conventions for multiplying matrices in the  auxiliary space, namely
 $(XY)_\nu^\mu = X_\lambda^\mu Y^\lambda_\nu  $.  We thus arrive finally
 to a label--independent expression for the transfer matrix:  $$t= {\rm
tr}_a \left( \cR _{aL} \cR _{aL-1} \cdots \cR _{a2} \cR _{a1} \right) \tag
mx25$$ Here, ${\rm tr}_a$ denotes the trace over the auxiliary space
$V_a$.

 After these preliminaries, we may ask whether two transfer matrices $t$
and $t'$, derived from two sets of Boltzman weights $\cR $ and $\cR '$,
do commute. Of course, $t$ and $t'$ must act on the same  quantum space
$V_1\otimes \cdots \otimes V_L$, but the auxiliary spaces for each of them
may be different. We multiply $t$ and $t'$  and for clarity we  label
their respective  auxiliary spaces as $V_a$ and $V_b$, even in the case
when these spaces are isomorphic:
  $$ t\,t' = {\rm tr}_{a\times b} \left( \cR _{aL} \cR '_{bL} \cdots \cR
_{a1} \cR '_{b1} \right) \tag$$ where ${\rm tr}_{a\times b}$ denotes the
trace on $V_a\otimes V_b$. Similarly, multiplying $t'$ and $t$ we get
  $$ t'\,t = {\rm tr}_{a\times b} \left( \cR '_{bL} \cR _{aL} \cdots \cR
'_{b1} \cR _{a1} \right) \tag$$ Hence $t$ commutes with $t'$ if and only
if there exists an invertible matrix $X_{ab}$ such that $$\cR '_{bi} \cR
_{ai} = X_{ab} \cR _{ai} \cR '_{bi} X^{-1}_{ab} \qquad \forall i=1,\ldots
,L \tag mx28$$ Indeed, using the cyclicity of the trace we find
$$\eqalignno{ t'\,t= {\rm tr}_{a\times b} \Bigl( X_{ab}\cR _{aL} \cR
'_{bL}  X^{-1}_{ab } X_{ab} \cR _{aL-1}  \cR '_{bL-1}  X^{-1}_{ab}\cdots &
\cr \cdots X_{ab} \cR _{a1} \cR '_{b1} X^{-1}_{ab }\Bigr) =t\,t' &&(mx29)
\cr} $$ Moreover, the matrix $X_{ab}$ may be interpreted as arising from
Boltzmann weights on the space $V_a\otimes V_b$: we shall call them $\cR
^{\prime\prime}_{ab}$. The integrability condition \(mx28) is the
Yang--Baxter equation in operator formalism:  $$\cR ^{\prime\prime}_{ab}
\cR _{ai} \cR '_{bi} = \cR '_{bi} \cR _{ai} \cR ^{\prime\prime} _{ab} \tag
mx210$$

    With some minor changes in notation, equation \(mx210) can be written
as  $$\cR _{12} \cR '_{13} \cR ^{\prime\prime}_{23} = \cR
^{\prime\prime}_{23} \cR '_{13} \cR _{12} \tag mx211$$ where $\cR _{12}$,
$ \cR '_{13}$ and $ \cR ^{\prime\prime}_{23}$ are Yang--Baxter matrices
acting on the spaces $V_1\otimes V_2$, $V_1\otimes V_3$ and $V_2\otimes
V_3$, respectively. In components, the  {operator Yang--Baxter equation}
\(mx211)   reads as follows:  $$ \sum _{j_1, j_2, j_3} \cR _{j_1 j_2}^{
k_1 k_2} \cR _{i_1 j_3}^{\prime j_1 k_3} \cR _{i_2 i_3}^{\prime\prime j_2
j_3} = \sum _{j_1, j_2, j_3} \cR _{j_2 j_3}^{ \prime\prime k_2 k_3} \cR
_{j_1 i_3}^{\prime k_1 j_3} \cR _{i_1 i_2}^{ j_1 j_2} \tag mx212$$

 Equation \(0) is the most general form of the Yang--Baxter equation for
vertex models, in the sense that the spaces $V_1$, $V_2$ and $V_3$ need
not be isomorphic. We shall not consider this possibility, but keep all
these vector spaces   two--dimensional, so that the $\cR $ operator is a
$4\times4$ matrix which, in the case of the six--vertex model, is  $$\cR
^{(6v)} (a,b,c) = \pmatrix { a &&&\cr &b&c&\cr &c&b&\cr &&&a\cr} \tag
mx213$$ If $\cR ^{(6v)}$ is to be invertible, then we must require
$a\not=0$ and $b\not= \pm c$.

  Taking now three six--vertex $\cR $--matrices $\cR =\cR ^{(6v)}
(a,b,c)$, $\cR '=\cR ^{(6v)} (a',b',c')$ and  $\cR ^{\prime\prime}=\cR
^{(6v)} (a^{\prime\prime},b^{\prime\prime},c^{\prime\prime})$, then the
Yang--Baxter equation holds provided  $$\Delta (a,b,c) =  \Delta
(a',b',c') =  \Delta (a^{\prime\prime},b^{\prime\prime},c^{\prime\prime})
\tag $$ in full agreement with equation \(mx172). The Yang--Baxter
equation captures completely the integrability of the
 {six--vertex model}, encoded in \(0).

  Expressing the weights $a$, $b$ and $c$ in terms of $u$, we find that
the Yang--Baxter matrix $\cR (u)=\cR ^{(6v)}(a(u), b(u) , c(u))$ satisfies
the Yang--Baxter equation \(mx211) in the form  $$\cR _{12}(u) \cR
_{13}(u') \cR _{23}(u^{\prime\prime}) = \cR _{23}(u^{\prime\prime}) \cR
_{13}(u') \cR _{12} (u)\tag $$ with $u^{\prime\prime}$ fixed in terms of
$u$ and $u'$. Now, on a sphere all points are equivalent in the sense that
any point can be mapped to any other one by a {conformal transformation}.
We may therefore choose the functions $a(u)$, $b(u)$ and $c(u)$ in such a
way that $u^{\prime\prime}$ is just $u'-u$. Then \(0) adopts the usual
additive form  $$\cR _{12}(u) \cR _{13}(u+v) \cR _{23}(v) = \cR _{23}(v)
\cR _{13}(u+v) \cR _{12} (u)\tag mx216$$
 valid for any complex $u$ and $v$.

The {monodromy matrix} $T(u)$ is defined in the same manner as the
transfer matrix, except that we do not trace over the first (or last, due
to periodic boundary conditions) horizontal states in \(mx24), that is to
say $$T(u) = \cR _{aL} \cR _{aL-1} \cdots \cR _{a2} \cR _{a1} \tag
monohuhu$$ The trace of the monodromy matrix on the  {auxiliary space} is
just the  {transfer matrix} $$t(u) = {\rm tr}_a T(u) \tag mx218$$ Using
$i,j,\ldots$ as labels in the auxiliary space $V_a$, we see that $T(u)$ is
in fact a matrix $T_i^j(u)$ of operator valued functions which act, in
this case, on the Hilbert space ${\cal  H}^{(L)} = V_1 \otimes \cdots
\otimes V_L$. These operators will be  represented graphically  as
$$T_i^j(u) =\qquad i\ {\quad \Big|\Big| \quad \over \Big|\Big|}\ j \tag
mx219$$ with the double line standing for the Hilbert space ${\cal
H}^{(L)}$. The characteristic feature of these operators is that they
satisfy an important set of quadratic relations reflecting their behavior
under monodromy:  $$ \cR _{ab}(u-v) \left( T_a(u) \otimes T_b(v) \right) =
\left( T_b(v) \otimes T_a(u) \right) \cR _{ab} (u-v) \tag mx220$$ This
equation constitutes the cornerstone of the  quantum inverse scattering
method; it is also at the origin of the quantum group. The subindices $a$
and $b$ are short-hand for the two auxiliary spaces $V_a$ and $V_b$ on
which the operators $T$ and $\cR $ act.    For extra clarity, we have
indulged in a notational redundance indicating the tensor product in \(0)
which  is taken over these auxiliary spaces, while the quantum indices
(not shown) are multiplied as ordinary matrix indices. The proof of \(0)
uses the Yang--Baxter equation \(mx216)  repeatedly and elucidates the
index interplay: $$ \eqalignno{ \cR _{ab}(u-v)  &\left( T_a(u) \otimes
T_b(v) \right)  =  \cr  &=  \cR _{ab}(u-v) \cR _{aL}(u) \cR _{bL}
(v)\cdots \cR _{a1}(u) \cR _{b1}(v)  \cr &=  \cR _{bL} (v) \cR _{aL}(u)
\cdots \cR _{b1}(v) \cR _{a1}(u)  \cR _{ab}(u-v) &()\cr &=
 \left( T_b(v) \otimes T_a(u) \right)
\cR _{ab} (u-v) \cr }    $$ For practical purposes, it is often convenient
to write equation \(+11) in components:
 $$ \sum_{j_1, j_2}  \cR _{j_1j_2}^{k_1k_2}(u-v)  T(u)_{i_1}^{j_1}  \;
 T(v)_{i_2}^{j_2}   =  \sum_{j_1, j_2}  T(v)_{j_2}^{k_2} \;
T(u)_{j_1}^{k_1}   \cR _{i_1i_2}^{j_1j_2} (u-v) \tag mx222$$ Given
\(mx220), it is an easy task to prove  the commutativity of the transfer
matrices, \ie $[{\rm tr}\, T_a(u) , {\rm tr}\, T_b(u) ]=0$.

 Equation \(mx222) has been derived for the {six--vertex model}, but it
can be taken as the starting point for the construction of integrable
vertex models, at least for those with the difference property. To this
end, we shall introduce the formal notion of a  Yang--Baxter algebra.

 A Yang--Baxter algebra $\cA$ consists of a couple $(\cR ,T)$, where $\cR
$ is an $n^2\times n^2$  invertible matrix and $T_i^j(u)$
($i,j\in\{1,\ldots,n\}; u\in\complex$) are the generators of $\cA$. They
must satisfy the quadratic relations \(mx222), whose consistency implies
the Yang--Baxter equation  \(mx216) for   $\cR (u)$. The entries of the
matrix $\cR (u)$ play the role of structure constants   of the
algebra $\cA$.    This is quite analogous to a Lie algebra, or better yet
to its universal enveloping algebra, which is also defined in terms of a
set of generators and structure constants. Following this analogy, the
Yang--Baxter relation plays the role of the {Jacobi identity}: they both
reflect the associativity of the corresponding algebras.

  An important property of Yang--Baxter algebras is their ``addition
law'', called co-product or co-multiplication    $\Delta$, which maps the
algebra $\cA$ into the tensor product $\cA\otimes \cA$ while preserving
the algebraic relations of $\cA$: $$\eqalign{ \Delta: \ & \cA  \to \cA
\otimes \cA \cr & T_i^j(u) \mapsto \sum_k T_i^k(u) \otimes T_k^j (u) \cr }
\tag mx223$$ The diagrammatic representation of the co-product follows
from \(mx219): $$ \Delta \left( i\ {\quad \Big|\Big| \quad \over
\Big|\Big|}\ j \right) =  \sum_k \  i\ {\quad \Big|\Big| \quad \over
\Big|\Big|} k{\quad \Big|\Big| \quad \over \Big|\Big|}\ j \tag $$ It is
left as an exercise for the reader to check that $\Delta T_i^j$ satisfy
the same relations as $T_i^j$ in \(mx222). The algebra $\cA$ has thus both
a multiplication and a co-multiplication; $\cA$ is called a  {bi-algebra}.

The definition of  a Yang--Baxter algebra just provided is general,  and
it can be applied to any $\cR $--matrix satisfying the Yang--Baxter
relation. We  confine ourselves once more to the Yang--Baxter algebra
constructed from the $\cR $--matrix of the six--vertex model.

We represent the four generators $T_i^j$ of $\cA^{(6v)}$ as  operators
acting on a Hilbert space $\cal H$. We shall name them as follows, for
convenience: $$\eqalign{ T_0^0 (u) = A(u) \quad,&\quad T_1^0 (u) = B(u)
\cr  T_0^1 (u) = C(u) \quad,&\quad T_1^1 (u) = D(u) \cr}\tag mx224$$ Just
as the structure constants  of Lie algebras provide a representation of
the algebra (the adjoint), the $\cR ^{(6v)}$ matrix provides a
representation of  $\cA^{(6v)}$ of dimension two under the identification
$$\left( T_i^j(u) \right) _\ell^k   \quad = \quad \cR _{i\ell}^{jk} (u)
\tag$$  or explicitly $$\eqalignno{ & A(u)   =\pmatrix{ a(u) & 0 \cr 0&
b(u) \cr}  = {a+b\over 2}\;\one + {a-b\over2} \;\sigma_3 \cr
 & B(u)  = \pmatrix{ 0 & 0 \cr c(u)&  0 \cr}   \quad = \quad c\;
\sigma^- \cr & C(u)  = \pmatrix{ 0 & c(u) \cr0 &  0 \cr}   \quad = \quad
c\; \sigma^+ &(mx225)\cr & D(u) = \pmatrix{ b(u) & 0 \cr 0&  a(u) \cr}  =
{a+b\over 2}\;\one - {a-b\over2} \;\sigma_3\cr } $$

Equations \(0) yield what we might call the spin--$\12$ representation of
the algebra \(mx222). This nomenclature is appropriate since $C(u)$ and
$B(u)$ act as raising and lowering operators, respectively, while $A(u)$
and $D(u)$ span the  {Cartan subalgebra}  of $SU(2)$. Using now the
{bi-algebra} structure of $\cA^{(6v)}$ defined by the co-multiplication
\(mx223), we may obtain a representation of  $\cA^{(6v)}$ on the space
${\cal H}^{(L)} = \bigotimes^L V_{\12}$. In particular, for $L=2$ we get
$$\eqalign{ & \Delta (A(u)) = A(u) \otimes A(u) + C(u) \otimes B(u) \cr
 & \Delta (B(u)) = B(u) \otimes A(u) + D(u) \otimes B(u) \cr
 & \Delta (C(u)) = C(u) \otimes D(u) + A(u) \otimes C(u) \cr & \Delta
(D(u)) = D(u) \otimes D(u) + B(u) \otimes C(u) \cr}\tag marmiamai$$ It is
an easy exercise to check  that $\Delta C$ annihilates the  {reference
state} $\ket\Omega \equiv \ket{00} = \ket{\Uparrow \Uparrow} $:  $$\Delta
C(u) \ket{00} = 0 \tag$$

 Under this interpretation of   $B(u)$ and $C(u)$ as creation and
annihilation operators, it follows   that $\Delta B(u)$ acting on the
reference state $\ket\Omega$ yields a  state in the sector with the number
of spins down equal to one. We can rewrite this state as  $$\Delta B(u)
\ket\Omega = \ket{\Psi_1} = \sum_x f(x) \ket{x}  = f(1)  \ket{10}  + f(2)
\ket{01} \tag mx229$$ with $$  f(1) = c(u) a(u) \qquad, \qquad  f(2)  =
b(u) c(u)   \tag$$ Comparing    \(0) with   \(mx133),   we deduce the
relation between Boltzmann weights and quasi-momenta:
$${b(u) \over a(u) } = \e^{ik} \tag fal233bis$$

 This method for lowering spins (\ie creating 1's) from a reference state
by means of the $B$ operators can be extended to a lattice with $L>2$
sites. To do so we recall the definition \(mx224)  of the operator $B$ as
the entry $T_1^0$ of the  {monodromy matrix}; thanks to the co-product
\(mx223), it can be made to  act  on the space ${\cal H} = \otimes^L
V_{\12}$:  $$B(u) = \Delta^{ L-1 } \left( T_1^0 (u) \right)  \tag $$ where
$$\Delta ^{ L-1 }: \cA \to \overbrace{\cA \otimes \cdots  \otimes
\cA}^{L\;\rm times} \tag$$ is the associative generalization of \(mx223),
$\Delta^{L-1}= (\one\otimes \Delta)\Delta^{L-2}$  with $L\ge2$. Hence a
state with $M$ spins down can be built as follows: $$\ket{\Psi_M} =  \prod
 _{i=1}^M B(u_i)  \ket{00\cdots0}  =  \prod _{i=1}^M B(u_i) \ket\Omega \tag
mx231$$

The states \(0) are called algebraic Bethe ansatz states [``algebraic'' in
contrast with the ``co-ordinate'' description \(genricus)], and constitute
a very good starting point for solving the eigenvalue problem of the
{transfer matrix}. In order to show this, let us work out   more
explicitly the relations satisfied by the generators of the  six--vertex
Yang--Baxter algebra.

{}From \(mx213) and \(mx222) we obtain, for arbitrary $u$ and  $v$,
$$\eqalignno{ & B(u) B(v)  = B(v) B(u) & (mx232 a) \cr
 &A(u) B(v) = {a(v-u) \over b(v-u)} B(v) A(u) - {c(v-u) \over b(v-u)} B(u)
A(v) & (mx232  b) \cr &D(u) B(v) = {a(u-v) \over b(u-v)} B(v) D(u) -
{c(u-v) \over b(u-v)} B(u) D(v) & (mx232  c) \cr &C(u) B(v) - B(v) C(u) =
{c(u-v) \over b(u-v)} \left( A(v) D(u) - A(u) D(v) \right) \quad & (mx232
d) \cr} $$ Equation \(mx232 a) implies that the algebraic Bethe ansatz
state \(mx231) is independent of the ordering in which the $B$ operators
are multiplied.

 The transfer matrix of the six--vertex model can be written from
equations \(mx218) and \(mx224) as
$$t^{(6v)} (u) = {\rm tr}_a T^{(6v)}(u) = A(u) + D(u) \tag $$
 Therefore, the problem of diagonalizing the transfer matrix \(0) in the
algebraic Bethe ansatz basis \(mx231) amounts to finding a choice of the
parameters $\{u_i, i=1,\ldots,M\}$ such that $$ \eqalign{ t^{(6v)}(u)
\ket{\Psi_M} \ &= \left[ A(u) + B(u) \right] \prod_{i=1}^M B(u_i)
\ket\Omega \cr &=  \Lambda_M (u; \{u_i\})   \prod_{i=1}^M B(u_i)
\ket\Omega \cr}\tag$$ The advantage of using the  algebraic Bethe ansatz
states   is that the whole computation involved in \(0) reduces to a
systematic use of the commutation relations \(mx232), in addition to the
obvious relations $$ A(u) \ket\Omega = a(u)^L \ket\Omega \qquad ,\qquad
D(u) \ket\Omega = b(u)^L \ket\Omega  \tag $$ Indeed, using \(mx232 b),
\(mx232 c) and \(0), we find  $$ \eqalign{  &\left( A(u) + D(u)
\right)     \prod_{i=1}^M B(u_i)  \ket\Omega  =   {\rm unwanted\
terms}\;+  \cr & + \left[ a^L(u) \prod _{i=1}^M {a(u_i-u) \over b(u_i-u) }
+ b^L(u)  \prod _{i=1}^M {a(u-u_i) \over b(u-u_i) } \right] \prod_{i=1}^M
B(u_i)  \ket\Omega  \cr}\tag ejercitovv$$ The first term of the
right--hand side of this equation gives us the eigenvalue of the transfer
matrix:  $$ \Lambda_M (u;\{u_i\})  = a^L(u) \prod _{i=1}^M {a(u_i-u) \over
b(u_i-u) } + b^L(u)  \prod _{i=1}^M {a(u-u_i) \over b(u-u_i) } \tag
mx237$$ From the second summands in \(mx232 b) and \(mx232 c), however,
we also obtain terms which are not of the desirable form $\prod_{i=1}^M
B(u_i)  \ket\Omega $.  If these terms are present, the algebraic  Bethe
ansatz does not work. The   {unwanted terms} actually cancel, for the
six--vertex model, under a judicious choice of the $u_i$ parameters.  The
condition that the parameters $u_i$ should satisfy to guarantee the
cancellation of the unwanted terms is precisely  the  {Bethe equations},
written in the form $$\left( {a(u_i) \over b(u_i) } \right)^L =
\prod_{{j=1\atop j\not=i}}^M  { a(u_i-u_j) b(u_j-u_i)  \over a(u_j-u_i)
b(u_i-u_j) }  \tag mx238$$

Equations \(mx237) and \(mx238) are the final outcome of the
diagonalization of the transfer matrix through the algebraic Bethe ansatz,
which we now can compare with   equations \(mx170) and \(mx171). Matching
the eigenvalues \(mx170) and \(mx237) yields  $$\eqalignno{ &{a(u_i -u)
\over b(u_i-u) } = P(k_i) = { a(u) b(u) + \left( c^2(u) -b^2(u) \right)
\e^{-ik_i(u_i)} \over a^2(u) -a(u) b(u) \e^{-ik_i(u_i)} } &(mx239 a)\cr
&{a(u -u_i) \over b(u-u_i) } = Q(k_i) = { a^2(u) -c^2(u) -a(u) b(u)
\e^{-ik_i(u_i)} \over a(u) b(u) -b^2(u) \e^{-ik_i(u_i)} } &(mx239 b)\cr}
$$ Choosing $u=0$ in \(0 a) and using the fact that $a(0)=c(0)\not=0$,
$b(0)=0$ (see the explicit  parametrization below),  we get  $${b(u_i)
\over a(u_i)} = \e^{ik_i(u_i)} \tag mx240$$ Hence the comparison between
\(mx238) and \(mx171) ends up producing  $$\hat{S}_{ji} = { a(u_j-u_i)
b(u_i-u_j)  \over a(u_i-u_j) b(u_j-u_i) } =\ -\ {  1-2\Delta \e^{ik_i} +
\e^{i(k_i+k_j)} \over   1-2\Delta \e^{ik_j} + \e^{i(k_i+k_j)} } \tag
mx241$$ which confirms the result \(fal233bis).  Equations \(+12), \(+11)
and \(0) provide the map between the quasi-momenta $k_i$ and the
uniformization variables $u_i$ used in the algebraic Bethe ansatz
construction.

Let us use the following   uniformization of the Boltzmann weights of the
{six--vertex model}:  $$\eqalign{ & a(u) = \cR _{00}^{00} (u) = \cR
_{11}^{11} (u)  = \sinh (u+i\gamma) \cr  & b(u) = \cR _{10}^{10} (u) = \cR
_{01}^{01} (u)  = \sinh u \cr  & c(u) = \cR _{01}^{10} (u) = \cR
_{10}^{01} (u)  =i \sin \gamma \cr }\tag mx242$$ where the parameter
$\gamma$ is related to the  anisotropy  $\Delta$ by the relation  $\Delta
 =\cos \gamma $.   Using now the map  $$\e^{ik_j} = {\sinh u_j \over \sinh
(u_j+i\gamma) } \tag $$ it is easy to check that equations \(mx239) and
\(mx241) are satisfied with the six--vertex $\cR$--matrix \(+11), and that
the Bethe equations can be written as  $$\left( { \sinh (u_j+i\gamma)
\over \sinh u_j} \right)^L = \prod_{{k=1\atop k\not=j}}^M {\sinh (u_j-u_k
+i\gamma ) \over \sinh (u_j -u_k -i\gamma) } \tag mrmx251bb$$

 We may calculate the energy of the Bethe ansatz state \(mx231)   from the
eigenvalue of the transfer matrix \(mx237)  if we recall that the
hamiltonian is defined as  $$\eqalignno{  H\;&= i {\partial\over\partial
u} \log \left( t(u) \over a(u)^L \right) \Bigg|_{u=0}  &(hahayyy)\cr&=
{1\over 2\sin \gamma} \sum_{j=1}^L \left[ \sigma_j^x \sigma_{j+1}^x +
\sigma_j^y \sigma_{j+1}^y +  \cos \gamma \left( \sigma_j^z \sigma_{j+1}^z
-1 \right) \right] \cr}$$  Indeed, carrying out the computation
explicitly we find
$$ \eqalignno { E_M\left( \{u_j\} \right)\;& = i  {\partial\over\partial
u} \log \left( \Lambda_M  \left(u,\{u_j\} \right)  \over a(u)^L
\right)\Bigg|_{u=0} \cr &= - \sum_{j=1}^M {\sin \gamma \over \sinh u_j
  \,\sinh (u_j+i\gamma) } &()\cr  } $$  Similarly, the total momentum of
the same Bethe state is
  $$  P_M  \left( \{u_j\}\right)   = i \log \left( t(u) \over a(u)^L
\right) \Bigg|_{u=0} = i \sum_{j=1}^M \log {\sinh \left( u_j + i\gamma
\right) \over \sinh u_j } \tag  $$

 Equations \(mx239)--\(mx241)     explain the integrability content of the
Bethe equations of the six--vertex model, which is codified in the
Yang--Baxter equation.   The factorization properties for $S$--matrices
simply reflect the consistency of the Yang--Baxter algebra. We have thus
reinterpreted the factorization properties as integrability of the
six--vertex model.

\subchap{Physical spectrum of the Heisenberg spin chain}
  Let us abandon formalism for a while and come back to the simple system
of the one--dimensional  antiferromagnetic  Heisenberg model (the  {$XXX$
model})   $$H=J \sum_{i=1}^L \left( \vec{\sigma}_i  \cdot \vec{\sigma
}_{i+1} -1\right)   \qquad J>0\tag cg311$$ Since $J$ is positive,
neighboring spins tend to align antiparallel. If $J$ was negative, then it
would be favored for all spins to align in the same direction, and we
would be in the ferrromagnetic phase.

The first issue we should address about a one--dimensional spin system
like \(cg311)  is whether a particle interpretation  exists. By this, we
mean whether it is possible to define a Fock representation of the Hilbert
space of the model, such that the  vacuum $\ket0$ of the Fock space $\cal
F$ corresponds to the  ground state and the many--particle states to the
low lying excitations. A particle interpretation is readily available if
we establish the correspondence   $$  {\cal H}^{\ell.\ell.}_\infty \to
\ket0 \oplus {\cal H}_1 \oplus {\cal H}_2 \oplus \cdots   ={\cal F}\tag
cg302$$ where ${\cal H}^{\ell.\ell.} _\infty$ represents the Hilbert space
of the low lying excitations of the  hamiltonian \(cg311) in the limit
$L\to\infty$, and  ${\cal H}_n$ is the Hilbert space of    $n$ elementary
excitations. The Fock vacuum     $\ket0$ represents the
{antiferromagnetic ground state}.

 The two crucial questions about the  {elementary excitations} are

\list{{\it ii}}

\item{{\it i}}The {dispersion relation} $\epsilon(k)$, from which we   get
information about the existence of a  {mass gap}.  If the mass gap is
zero, then the theory may correspond in the continuum limit to a massless
or critical field theory, \ie to a conformal field theory.

 \item{{\it ii}}The internal quantum numbers (spin) of the elementary
excitations.

\endlist

 To get some flavor for why the answer to these two questions is so
difficult,  it is necessary to reflect for a moment on the richness of the
antiferromagnetic vacuum. Recall     that to solve the model \(cg311) we
must diagonalize the hamiltonian in the basis of spin waves. A state with
$M$ spin waves is   gotten by flipping $M$ spins down from the reference
state with all spins up. For the trivial case of only one  {spin wave}  in
a periodic  chain, the dispersion relation for the spin wave of the $XXX$
model is given by  $$E(k) = 4J ( \cos k-1) \tag cg315$$ and thus the
different physical behavior of  the ferromagnetic phase and the
antiferromagnetic one  can be easily distinguished.

In the  {ferromagnetic} regime, the coupling constant $J$ is negative and
the energy of the  spin wave is positive, so the Bethe  reference state
 coincides with the ground state of minimal energy; this is the ordered
phase, and spins tend to align.  The solution to the physical problem is
relatively straightforward.

 The antiferromagnetic regime, with a positive coupling $J>0$, is
trickier. The energy of the spin wave is negative, and flipping one spin
down is energetically favored over keeping all spins up.  The Bethe
reference state has nothing to do with the  ground state, which we
expect  to be a singlet of the global $SU(2)$ symmetry, in fact a state
with $S^z_{\rm total}=0$. To get such a state in our picture, we need a
``condensate'' of spin waves. This physical intuition, together with the
fact that the energy of the spin wave for $J>0$ is negative, can be
combined thanks to   the concept of a  {Dirac sea}. Identifying the vacuum
as the Dirac sea filled up to the  {Fermi surface}, the elementary
excitations will be thought of as holes in the Dirac sea.  The
 integrability of the model, which  amounts to the factorizability of the
scattering matrix for spin waves, allows us to construct the sea starting
from the Bethe equations.

 For the  isotropic Heisenberg model \(cg311)  in the antiferromagnetic
phase $J>0$, the  two questions above were pretty much solved  by Faddeev
and Takhtadjan in the early eighties. The dispersion relation for the low
lying excitations turns out to be of the form  $$\epsilon(k)=2\pi J \,\sin
k \qquad 0 \le k \le \pi \tag mr33$$ which means that the system has no
mass gap. This is consistent with a continuum limit described by a free
massless scalar field. The surprising result has to do  with   the spin of
the low lying excitations.  Since an excitation corresponds to flipping
one local spin up into a spin down, with a net change of one unit of
angular momentum, you might have guessed from   the one--particle
hamiltonian  that the  elementary excitations would have spin one.
Instead, it turns out that the particle--like excitations over the
antiferromagnetic Dirac sea have spin $\12$. The Fock space of the model
is thus, for a chain with an even number of sites,   $${\cal F} =
\matrix{\infty\cr\bigoplus\cr n=0\cr } \int_0^\pi \cdots \int_0^\pi dk_1
\cdots dk_{2n} \otimes^{2n} \complex^2   \tag mr34$$ where the
integrations run over the possible values of the momenta and $\complex^2$
represents the internal spin--$\12$ space of dimension two.  Proper
symmetrization of the states in \(0) must   be taken into account as well.
The excitations come in pairs [whence the $2n$ in \(0)], for otherwise the
total spin of a chain with an even number of sites would not be an
integer.  Let us stress that the internal quantum numbers of the
elementary excitations are a completely unexpected collective result,
impossible to predict {\sl a priori}. This  is   the motivation for
invoking particles with strange statistics (anyons) in attempts to
understand high temperature superconductors.

 \subchap{Yang--Baxter algebras and braid groups} The basic relation
studied two section ago  is the Yang--Baxter equation \(mx216).  In
components, it reads as   $$\eqalign{  \sum _{j_1, j_2, j_3} \cR _{j_1
j_2}^{ k_1 k_2}&(u) \cR _{i_1 j_3}^{ j_1 k_3} (u+v) \cR _{i_2 i_3}^{ j_2
j_3} (v)=\cr &= \sum _{j_1, j_2, j_3} \cR _{j_2 j_3}^{  k_2 k_3} (v) \cR
_{j_1 i_3}^{k_1 j_3} (u+v)  \cR _{i_1 i_2}^{ j_1 j_2} (u) \cr} \tag
mx251$$ where all the indices run from 1 to $n={\rm dim}\,V$ ($n=2$ for
the  six--vertex model). An interesting way to write this equation calls
for the permuted $R$ matrix,  $$R  = P \cR : V_1\otimes V_2 \to V_2
\otimes V_1 \tag$$ where $P$ is the {permutation map}  $$\eqalign{ P:
V_1\otimes V_2 \ & \to V_2\otimes V_1 \cr e_i^{(1)} \otimes e_j^{(2)}
&\; \mapsto e_j^{(2)} \otimes e_i^{(1)} \cr} \tag mx252$$ with
  $\{e_r^{(i)}, r=1,\ldots,n\}$   a basis of $V_i$. The relation between
the entries of $R $ and $\cR$ is straightforward: $$R  =P\cR \quad \iff
\quad R ^{k\ell}_{ij} = \cR_{ij}^{\ell k} \tag mx253$$ With the help of the
permuted $R $--matrix, the Yang--Baxter equation \(mx251) can be written
as  $$\eqalign{ \left( \one \otimes R (u) \right) &\left( R (u+v)  \otimes
\one\right)  \left( \one \otimes R (v) \right) =\cr &=  \left( R (v)
\otimes \one\right) \left( \one \otimes R (u+v) \right)  \left( R (u)
\otimes \one\right) \cr} \tag mx254$$ Every operator in parentheses acts on
the space $V\otimes V \otimes V$:  $$\eqalign{ &\left( R (u)  \otimes
\one\right)  e_{i_1}\otimes   e_{i_2}\otimes   e_{i_3}  =   R  _{i_1
i_2}^{ j_2 j_1} (u) \;  e_{j_2}\otimes   e_{j_1}\otimes   e_{i_3} \cr
 &\left(  \one\otimes R  (u)  \right)  e_{i_1}\otimes   e_{i_2}\otimes
e_{i_3}   =   R  _{i_2 i_3}^{ j_3 j_2} (u) \;  e_{i_1}\otimes
e_{j_3}\otimes   e_{j_2} \cr }\tag$$
 Note that $R$ is very close to a factorizable $S$--matrix.

 The reason for writing the Yang--Baxter equation in the form \(+11) comes
from its relation to the  {braid group}   $B_L$ on $L$ strands, which is
generated by $L-1$ elements $\sigma_i$ ($i=1,\ldots,L-1$) subject to the
relations $$\eqalignno{ &\sigma_i \sigma_{i+1} \sigma_i =  \sigma_{i+1}
 \sigma_i\sigma_{i+1} &(mx256 a)\cr  &\sigma_i \sigma_j = \sigma_j \sigma_i
\qquad |i-j|\ge2 &(mx256 b)\cr & \sigma_i \sigma_i^{-1} = \sigma_i^{-1}
\sigma_i  =\one &(mx256 c) \cr } $$ The generator $\sigma_i$ braids the
$i$--th strand under the $(i+1)$--th strand, whereas $\sigma_i^{-1}$
effects the inverse braiding, \ie it takes the $i$--th strand over the
$(i+1)$--th strand.

 Trying to make \(mx254) look more like \(mx256), we define the operators
$ R  _i(u)$ (for $i=1,\ldots,L-1$) on $ \bigotimes_{i=1}^L V_i$, which act
on the spaces $V_i \otimes V_{i+1}$ as $ R  (u)$ and as the identity
elsewhere: $$ R  _i(u) = \one\otimes\cdots\otimes\one\otimes \matrix{
(i,i+1)\cr R   (u) \cr\cr} \otimes\one\otimes \cdots \otimes\one \tag$$
Then equation \(mx254) becomes  $$ R  _{i+1}(u)  R  _{i}(u+v)  R
_{i+1}(v)  =   R  _{i}(v) R  _{i+1}(u+v)  R  _{i}(u) \tag$$ while
obviously $$ R  _{i}(u)  R  _{j}(v) =  R  _{j}(v) R  _{i}(u) \qquad
|i-j|>1 \tag$$ The identification of the Yang--Baxter equation in the form
\(+11) with the braid group relation \(mx256 a) cannot be realized yet due
to the presence of the  rapidity variable $u$. This should be no problem,
however, in a situation where $u=v=u+v $ which  has two solutions:

\list{{\it ii}}

 \item{{\it i}} $u=v=0$,

\item{{\it ii}} $u=v$, $|u|=\infty$.

\endlist

 The first solution is trivial, since from \(mx242) and \(mx253) we get
merely  $$  R  (u=0)  = i \sin \gamma\ \one  \tag   $$ And thus $\cR$ is
just proportional to a permutation $P$. Solution ({\it ii}) is known as
the  {braid limit}. Up to constant factors,  $$ \lim_{u\to \pm \infty}
\e^{-|u|}  R  (u) \sim P \exp\left[(\pm i \gamma/2)  \sigma^z\otimes
\sigma^z\right] \tag mx262$$ where we have assumed $u$ real.  This limit
provides us with a representation of the  braid group in terms of,
essentially, permutations. The permutation group of $L$ elements ${\cal
S}_L$ satisfies the same defining relations as the braid group $B_L$,
except for the crucial difference that the square of a transposition is
the identity, and therefore one cannot distinguish   overcrossings  from
undercrossings.  A good representation of the braid group should be able
to distinguish between $\sigma$ and $\sigma^{-1}$.   In the limit $u\to
+\infty$, the Boltzmann weights behave as  $$  a(u) \to \12 \e^u
\e^{i\gamma}  \quad ,\qquad  b(u) \to \12 \e^u    \quad ,\qquad  c(u) = i
\sin \gamma  \tag$$ Hence the information contained in the weight $c(u)$
is washed out in the limit, which accounts for the ``triviality'' of the
result \(+11). In order not to lose information in the limits $u\to \pm
\infty$, we    perform a   $u$--dependent rescaling of the basis elements,
\ie a $u$--dependent ``diagonal'' change of basis $\tilde{e}_r(u)   =
f_r(u) e_r(u)$ ($r=1,\ldots,n)$. Recalling that the definition of $ R $
is  $$ R  \left( e_{r_1} (u_1) \otimes  e_{r_2} (u_2) \right) =   R _{r_1
r_2} ^{r'_2r'_1}  (u_1-u_2)   e_{r'_2} (u_2) \otimes e_{r'_1} (u_1) \tag
mx264$$ we deduce that the $ R $--matrix in the new basis $\tilde{e}_r$ is
given by  $${\tilde R}_{r_1 r_2} ^{r'_2r'_1} (u_1, u_2) = {f_{r_1} (u_1)
f_{r_2} (u_2)  \over  f_{r'_1} (u_1) f_{r'_2} (u_2) } { R}_{r_1 r_2}
^{r'_2r'_1} (u_1- u_2)  \tag$$ The trick is to preserve the  {difference
property} of the  {$R$--matrix} under this change of basis, and thereby
fix the scaling functions $f_r(u)$.     Indeed, if ${\tilde R} (u_1, u_2)$
is to still depend only on the difference $u_1-u_2$, the functions in the
change of basis must be $f_r(u) = \e^{\alpha  u r}  $ Since  ${\tilde R}
$   conserves the total quantum number, that is  $ {\tilde R}_{r_1 r_2}
^{r'_2r'_1} (u_1, u_2) =0 $  unless $ r_1 + r_2 = r'_1 + r'_2  $, we may
write the rescaled $\tilde{R}$ matrix explicitly as   $${\tilde R}_{r_1
r_2} ^{r'_2r'_1} (u_1- u_2, \alpha) = \e^{\alpha (u_1-u_2) (r_1-r'_1)} {
R}_{r_1 r_2} ^{r'_2r'_1} (u_1- u_2)  \tag$$
  We  take the  braid limit of \(0) at a special value of $\alpha$:
 $${R}  \equiv 2\e^{-i\gamma/2} \lim_{u\to+\infty} \e^{-u}   {\tilde R} (u,
\alpha=1)  \tag mx268$$ obtaining
$$\eqalign{ &  R _{00}^{00} =  R _{11}^{11} =  \e^{i\gamma/2} \cr
&  R _{10}^{01} =  R ^{10}_{01} =  \e^{-i\gamma/2} \cr
&  R _{10}^{10} = \e^{-i\gamma/2}  \left( \e^{i\gamma}
-\e^{-i\gamma} \right) \cr
&  R _{01}^{01} =  0 \cr} \tag mx269$$ or, in matrix form,
 $$ R =\bordermatrix{ & 00 & 01 & 10 & 11 \cr 00 & q^\12 & 0 & 0 & 0 \cr
01 & 0 & 0 & q^{-\12} & 0 \cr 10 & 0 & q^{-\12} & q^{-\12} \left( q-q^{-1}
\right) & 0 \cr  11 & 0 & 0 & 0 & q^\12 \cr} \tag$$ with $$q=\e^{i\gamma}
\tag$$ These $R$--matrices, first derived by Jimbo, satisfy the
Yang--Baxter relation \(mx254) without {spectral parameter}, and   appear
often in the literature.   It can be checked explicitly that $( R )^2 \ne
\one$ for $\gamma\ne0$, so that we have indeed obtained a genuine
representation of the  braid group. In the isotropic case ($\gamma=0$) we
fall back to the previous result \(mx262). The inverse matrix $ R ^{-1}$
can be obtained as the other real infinite limit of the same rescaled $ R
$: $$ R ^{-1} = -2 \e^{i\gamma/2} \lim_{u\to-\infty} \e^u   R  (u,
\alpha=1) \tag$$

\subchap{Yang--Baxter algebras and quantum groups}
In the new basis $\{\tilde{e}_r\}$, the {$\cR $--matrix} (without the
 permutation in $ R $) and the  {monodromy matrix} $T$ are related to those
 in the basis $\{e_r\}$ by the equations  $$\eqalignno{ & \tilde{\cR
}_{i_1i_2}^{j_1j_2} (u) = \e^{u(i_1-j_1)} {\cR }_{i_1i_2}^{j_1j_2} (u) &
(mx272 a) \cr &\tilde{T} (u)_i^j = \e^{u(i-j)} {T} (u)_i^j & (mx272 b)
\cr}$$ The reader  may check that $\tilde{\cR }(u)$ and $\tilde{T}(u)$ do
satisfy indeed equation \(mx222).

 Now just like we took the limit $u\to\pm\infty$ of the matrix $\tilde{\cR
}(u)$, we may take the limit of the monodromy matrix $\tilde{T}(u)$. To
get a feeling for what this limit may yield, let us consider the
spin--$\12$ representation  given by equations \(mx225): $$\eqalignno{
\lim_{u\to+\infty}\tilde{A}(u) \;& = \lim_{u\to+\infty} \12\e^u
\e^{i\gamma/2} \pmatrix{\e^{i\gamma/2} &\cr& \e^{-i\gamma/2} \cr} = \12
\e^u q^{1/2} q^{S^z} \cr   \lim_{u\to-\infty} \tilde{A}(u) \;&=
\lim_{u\to-\infty} -\12\e^{-u} \e^{-i\gamma/2} \pmatrix{\e^{-i\gamma/2}
&\cr& \e^{i\gamma/2} \cr} &() \cr &= -\12 \e^{-u} q^{-1/2} q^{-S^z} \cr}
$$ where $S^z = \12 \sigma^z  $ is  the Cartan generator in the
spin--$\12$ representation of {$SU(2)$} and we recall that
$q=\e^{i\gamma}  $.   The limits $u\to\pm\infty$ of ${\tilde B}(u)$,
${\tilde C}(u)$ and ${\tilde D}(u)$ can be evaluated similarly, whereby
the braid limits  of the monodromy matrix $\tilde{T}(u)$ are
$$\eqalignno{   T_+ \;& \equiv  2q^{-1/2} \lim_{u\to+\infty} \e^{-u}
\pmatrix{  \tilde{T}_0^0 &  \tilde{T}_0^1 \cr \tilde{T}_1^0 &
\tilde{T}_1^1 \cr } \cr & = \pmatrix { q^{S^z} & 0 \cr q^{-1/2} (q-q^{-1})
S^- &   q^{-S^z}\cr}   & (mx275 a) \cr  T_- \;& \equiv -2q^{1/2}
\lim_{u\to-\infty} \e^{u} \pmatrix{  \tilde{T}_0^0 &  \tilde{T}_0^1 \cr
\tilde{T}_1^0 &  \tilde{T}_1^1 \cr } \cr & = \pmatrix { q^{-S^z} &
-q^{1/2} (q-q^{-1}) S^+ \cr 0&   q^{S^z}\cr}   & (mx275 b) \cr} $$
 where   $S^\pm= (\sigma^x \pm i \sigma^y)/2$ are the off--diagonal
generators of $SU(2)$ in the spin--$\12$ irrep.

 The fun starts when we take the various limits $u\to\pm\infty$,
$v\to\pm\infty$ in the $ \cR  {T} {T} =  {T}  {T}  {\cR }$ equation
\(mx220).    All the extra factors work out nicely so that the result is
$$\eqalign{& \cR_{12}  \left( T_+ \right)  _1   \left( T_+ \right)  _2 =
\left( T_+ \right)  _2   \left( T_+ \right)  _1  \cR_{12} \cr & \cR_{12}
\left( T_+ \right)  _1   \left( T_- \right)  _2 =  \left( T_- \right)
_2   \left( T_+ \right)  _1  \cR_{12} \cr & \cR_{12}^{-1}  \left( T_-
\right)  _1   \left( T_- \right)  _2 =  \left( T_- \right)  _2   \left(
 T_- \right)  _1  \cR_{12}^{-1}  \cr}\tag$$ with $\cR=P R $ [$ R $ given by
\(mx269)] and $T_\pm$ from \(+11).

 This system of equations appears rather complicated at first sight. They
are, in fact, equivalent to the following algebraic relations between $
{S^z}$, $S^+$ and $S^-$: $$\eqalignno{ & \left[{S^z} ,S^\pm \right]  =
\pm S^\pm  & (mx277 a) \cr & \left[ S^+  , S^-\right] = {
q^{2S^z}-q^{-2S^z}  \over q-q^{-1}}  & (mx277 b) \cr}$$ These are the
defining relations for the quantum group $U_q(s\ell(2))$, which is some
kind of deformation of the Lie algebra  $s\ell(2)$ with $q=\e^{i\gamma}$
acting as  {deformation parameter}.

The notation  {$U_q(s\ell(2))$} clarifies that the  quantum group consists
of all the formal powers and linear combinations of $S^+$, $S^-$ and
$S^z$, subject to the relations \(mx277). Traditionally,  $U(s\ell(2))$
denotes the  {universal enveloping algebra} of $s\ell(2)$, that is all the
formal powers and linear combinations of $S^\pm$ and $S^z$ modulo the
standard Lie algebra relations.  In the isotropic limit $\gamma\to0$, \ie
$q\to1$, we  recover from \(0) the usual $s\ell(2)$ algebra. The limit
$\gamma\to0$ is called classical, in the sense that the ``quantum'' group
$U_q(s\ell(2))$ becomes the ``classical'' universal enveloping   algebra
$U(s\ell(2))$. From the viewpoint of rigid nomenclature, it is perhaps
unfortunate that the classical limit of a quantum group is (the universal
enveloping algebra of) a   Lie algebra; beware of   misled distinctions
between quantum groups and quantum algebras!

 Finally, we may take the braid limits $u\to\pm\infty$  in the
co-multiplication rule \(mx223) to find the co-multiplication for the
generators of the quantum group $U_q(s\ell(2))$: $$\eqalignno{ & \Delta
(q^{S^z})  = q^{S^z} \otimes q^{S^z}& (mx278 a) \cr  & \Delta ({S^\pm})  =
S^\pm \otimes q^{S^z} + q^{-S^z}\otimes S^\pm & (mx278 b) \cr }$$ The
{co-multiplication} preserves the algebraic relations \(mx277), as can be
checked by using the fact that $\Delta$ is a homomorphism, that is to say
$\Delta(ab) = \Delta(a) \Delta(b)$. Note that the non--trivial addition
rule \(mx278 b) is consistent with the non--trivial commutator \(mx277 b),
and viceversa. Compare also with $\Delta B(u)$ in equation \(marmiamai).

 We have derived an interesting algebraic structure, the quantum
 group $U_q(s\ell(2))$, by letting the rapidities become infinite in the
Yang--Baxter elements $T_i^j(u)$. More precisely, the quadratic relations
between the  monodromy matrices ($\cR TT$ equations of the Yang--Baxter
algebra)  give us the defining relations of $U_q(s\ell(2))$, while the
co-multiplication of $\cA^{(6v)}$ implies that of $U_q(s\ell(2)) $.

 Let us return once again to the Yang--Baxter equation satisfied by the
$\cR $--matrix: $$\eqalign{  \sum _{j_1, j_2, j_3} \cR _{j_1 j_2}^{ k_1
k_2}&(u-v) \cR _{i_1 j_3}^{ j_1 k_3} (u) \cR _{i_2 i_3}^{ j_2 j_3} (v)=\cr
&= \sum _{j_1, j_2, j_3} \cR _{j_2 j_3}^{  k_2 k_3} (v) \cR _{j_1
i_3}^{k_1 j_3} (u)  \cR _{i_1 i_2}^{ j_1 j_2} (u-v) \cr} \tag mx279$$ The
$\cR TT$ equation is based on the identification of $T_i^j$ in the
representation of dimension two with the $\cR $--matrix itself: $$\left(
T_i^j(u) \right)_\alpha^\beta = \cR _{i\alpha}^{j\beta}  = \qquad  i \ {
\quad {\beta \atop \Big|\Big|} \quad \over {\Big|\Big|\atop \alpha } }\ j
 \tag$$ where $i$, $j$ are indices of the  {auxiliary space} (that is,
 labels for the elements of $\cA^{(6v)}$) and $\alpha$, $\beta$ are indices
 of the  {quantum space} (indicating the representation of $\cA^{(6v)}$).
We wish to emphasize that in all this construction, the $\cR $ matrix has
played a rather auxiliary role, as indeed its indices in $\cR TT=TT\cR $
are auxiliary:  we would like to see the $\cR $--matrix   playing a role
in quantum space as well.

 Let us take advantage of an interesting property satisfied by the $\cR
$--matrix of the  six--vertex model, the {parity symmetry}: $$\cR
_{i_1i_2}^{j_1j_2} (u) = \cR _{i_2 i_1} ^{j_2  j_1} (u) \tag mx281 $$  or
equivalently $$P\cR (u) P  = \cR (u) \tag mx282$$ with $P$ the
{permutation operator} \(mx252).  With the help of  equation \(mx281), we
may rewrite \(mx279) as  $$\eqalign{  \sum _{j_1, j_2, j_3} \cR _{j_1
j_2}^{ k_1 k_2}&(u-v) \cR _{ j_3i_1}^{  k_3j_1} (u) \cR _{ i_3i_2}^{ j_3
j_2} (v)=\cr &= \sum _{j_1, j_2, j_3} \cR _{ i_3j_1}^ { j_3k_1} (u)  \cR
_{ j_3j_2}^{  k_3 k_2} (v) \cR _{i_1 i_2}^{ j_1 j_2} (u-v) \cr} \tag
mx283$$ Note that in \(mx279) the space $V_1\otimes V_2$ is auxiliary and
$V_3$ quantum, whereas now $V_3$ is auxiliary but both $V_1$ and $V_2$ are
quantum. We have thus gotten $\cR $ into  quantum space. Using \(+13),
we   rewrite \(0) as  $$\cR (u-v)  \left( T_j^k (u) \otimes T_i^j (v)
\right) =  \left( T_i^j (u) \otimes T_j^k (v) \right)\cR (u-v) \tag
mx284$$ where the tensor product takes place in $V_1\otimes V_2$ and $\cR
(u-v) \in {\rm End}( V_1\otimes V_2) $.

 What are the consequences of  the $\cR TT$ equation  \(mx284) with $\cR$
in quantum space? First of all, for  consistency with the parity symmetry,
there must exist  some function $\rho (u)$ such that  $$\cR (u) \cR (-u) =
\rho(u)\rho(-u) \one \tag mx285$$   This can be derived by acting on both
sides of \(mx284) with the permutation operator $P$ and then using
\(mx282). Equation \(0) can also be derived from   the {unitarity}
condition \(unimxmra), namely $\cR(u) P \cR(u) P \sim \one$ for parity
invariant $\cR$--matrices, \ie satisfying \(mx284). For the $\cR$--matrix
of the six--vertex model,   $\rho(u)=a(u)$ in \(mx285).

 Letting $u=v$ in \(mx284) and knowing that $$\cR (0)= \rho(0) P \tag$$
we obtain  $$ T_i^j (u) \otimes T_j^k (u) = P  \left( T_j^k (u) \otimes
T_i^j (u) \right) P \tag$$ which establishes the equivalence between the
co-multiplication \(mx223) and its transpose.  More generally, when $u\ne
v$, equation \(mx284) establishes an equivalence between the  two ways of
co-multiplying arbitrary elements of $\cA$. Note that $u$ and $v$ are
labels of the representation spaces $V_1$ and $V_2$, respectively.  If,
recalling \(mx223),  we define    $$\eqalignno{ &\Delta_{u,v} \left( T_i^k
\right) =  T_i^j (u) \otimes T_j^k (v) &(mx287 a) \cr &\Delta'_{v,u}
\left( T_i^k \right) =  T_j^k (v) \otimes T_i^j (u) &(mx287 b)\cr }$$ we
can write \(mx284) as  $$\Delta_{u,v}\left( T_i^k \right) = \cR (u-v)
\Delta'_{v,u} \left( T_i^k \right) \cR ^{-1}(u-v) \tag mx288$$ This
equation means that $\cR (u-v)$ intertwines the two possible
{co-multiplication}s $\Delta_{u,v}$ and $\Delta'_{v,u}$. In Drinfeld's
definition of  {quantum group}s as quasi-triangular Hopf algebras,
equation \(0) is one of the basic postulates.

 To understand how the  {$\cR $--matrix} intertwines between a co-product
and its transpose, let us take a closer look at the  {braid limit}
$u\to\infty$ of equation \(mx284), or rather of its analog for the
rescaled $\tilde{\cR }(u)$ and $\tilde{T}(u)$ introduced in \(mx272),
namely   $$ \tilde{\cR } (v-u) \left( \tilde{T}_i^j (u) \otimes
\tilde{T}_j^k (v) \right)  =  \left( \tilde{T}_j^k (v) \otimes \tilde{T}
_i^j (u) \right) \tilde{\cR }(v-u) \tag$$  Using \(mx268) and \(mx275) we
find the braid limit of \(0), which is the braid limit of \(+11):
$$\Delta'(g) = \cR \Delta(g) \cR ^{-1}  \qquad g\in\left\{ q^{\pm S^z},
S^\pm\right\}\tag $$  Here, the $\cR$--matrix is  $\cR=P R $ with $ R $
the Jimbo matrix \(mx269), the co-product $\Delta(g)$ is given by
\(mx278), and the transposed  {co-product} $\Delta'$ is, explicitly,
$$\eqalign{ & \Delta' (q^{S^z})  = q^{S^z} \otimes q^{S^z}  \cr
  & \Delta' ({S^\pm})  = S^\pm \otimes q^{-S^z} + q^{S^z}\otimes S^\pm
\cr } \tag $$

 Equation \(+11) should really be written, to avoid confusion, as
$$\Delta_{\12\12}'(g) = \cR^{\12\12} \Delta_{\12\12}(g) \left(\cR
^{\12\12}\right) ^{-1} \tag mx291$$ where $\Delta_{\12\12}$ denotes the
restriction of $\Delta(g)$ to the irrep $\12\otimes\12$ of $U_q(s\ell(2))$
and the indices on $\cR$ remind us that we are using the representation of
$\cR$ on the vector space $\12\otimes\12$.  It is nevertheless worth
stressing that \(+12)   makes sense even at the level of the quantum group
$U_q(s\ell(2))$,  prior to the construction of its representations.  By
this we mean that $\cR$ in \(+11) can be viewed as an element of
$U_q(s\ell(2))\otimes U_q(s\ell(2)) $, rather than as a numerical matrix
as in \(0). For this reason, the matrix $\cR$ is called the ``universal
$R$--matrix'', and its existence guarantees that the co-multiplications
$\Delta$ and $\Delta'$ of  {$U_q(s\ell(2)) $} are equivalent at the purely
algebraic level.

\subchap{Affine quantum groups}
 We have found that the  integrability of the  {six--vertex model} in the
braid limit (when the dependence of the Boltzmann weights on the rapidity
drops out) is encoded in the quantum group $U_q(s\ell(2))$.  Motivated by
these results, we may ask ourselves whether a mathematical structure
similar to the quantum group could be associated with the
rapidity--dependent $R$--matrix solutions to the Yang--Baxter equation.

 The $s\ell(2)$ Lie algebra ($A_1$ in Cartan's classification) has three
Chevalley generators $E$, $F$ and $H$ with the following non--vanishing
commutators  $$\eqalign{ &  [E,F] = H \cr
 & [H,E]=2E \cr
 & [H,F]=-2F \cr}\tag mx292$$  The usual spin generators are related to
these by $$E=S^+ \;,\quad F=S^- \;,\quad H=2S^z \tag$$

 The affine extension of $A_1$, called  $A_1^{(1)}$ in Kac's
classification, has six Chevalley generators $E_i$, $F_i$ and $H_i$
($i=0,1$). Suppose we have an irreducible representation of   $A_1$. It
can be affinized, \ie promoted to an  {irreducible representation} of
$A_1^{(1)}$ through the identifications  $$\eqalign{ E_0 =\e^u F
\quad\qquad\qquad&\qquad E_1= \e^u E  \cr F_0 =\e^{-u} E
\qquad\qquad&\qquad  F_1 = \e^{-u} F \cr H_0=-H \quad\qquad\qquad&\qquad
H_1 = H \cr}    \tag mx293$$ where $x=\e^u$ is a complex  {affinization
parameter}.

 Different irreps of $A_1^{(1)}$ (of zero  {central extension}) may be
labelled by the  affine parameter $\e^u$ and the Casimir of the
corresponding representation of $A_1$. For example, the irreducible
$(\e^u,\12)$ representation of $A_1^{(1)}$ derives from the usual
spin--$\12$ irrep of $A_1$: $$ \eqalignno{ & E_0 = \pmatrix { 0 & 0\cr
\e^u & 0 \cr } \qquad\qquad \quad
 E_1 = \pmatrix { 0 & \e^u \cr 0 & 0 \cr } \cr
& F_0 =  \pmatrix { 0 & \e^{-u}\cr 0 & 0 \cr }\qquad\qquad  F_1 =
\pmatrix { 0 & 0\cr \e^{-u} & 0 \cr }  &(mx294) \cr
& H_0 =  \pmatrix { -1 & 0\cr 0 & 1 \cr }\qquad\qquad \ H_1 =
\pmatrix { 1 & 0\cr 0 & -1 \cr }  \cr} $$

 Let us turn to the quantum deformations of $A_1$ and $A_1^{(1)}$, which
we shall denote by
 {$U_q(A_1)$} and  $U_q(A_1^{(1)})$, respectively.
If we define the operator $K$ as
$$K=q^H\tag$$ we may rewrite equations
 \(mx277)   in terms of $E$, $F$ and $K$ as
 $$ \eqalignno{ & KE= q^2 EK \cr  & KF = q^{-2} FK &(mx295) \cr& [E,F] =
{K-K^{-1} \over q-q^{-1}} \cr} $$
 We shall denote by  $U_q(A_1)=U  _q(s\ell(2))$ the algebra generated by
$E$, $F$ and $H$ subject to \(mx277).

 Affinization of the spin--$\12$ irrep of $U_q(A_1)$ yields an irrep
$(\e^u,\12)$ of $U_q(A_1^{(1)})$:
 $$ \eqalignno{ & E_0 = \pmatrix { 0 & 0\cr \e^u & 0 \cr }
\qquad\qquad\quad
 E_1 = \pmatrix { 0 & \e^u\cr 0 & 0 \cr } \cr
& F_0 =  \pmatrix { 0 & \e^{-u}\cr 0 & 0 \cr }\qquad\qquad  F_1 =
\pmatrix { 0 & 0\cr \e^{-u} & 0 \cr }  &(mx296) \cr
& K_0 =  \pmatrix { q^{-1} & 0\cr 0 & q \cr }\qquad\qquad  K_1 =
\pmatrix { q & 0\cr 0 & q^{-1} \cr }  \cr}   $$
 which is   the same as representation \(mx294) of the classical group
$A_1^{(1)}$   provided we take into account   relation \(+12).  For the
fundamental irrep, as the doublet of $s\ell(2)$, it is always true that
the classical and quantum representations coincide.

 We expect that the irrep ($\e^u,\12$) of $U_q(A_1^{(1)})$ should be
intimately related to the spin--$\12$ representation \(mx225) of the
generators $A(u)$, $B(u)$, $C(u)$ and $D(u)$ of $\cA^{(6v)}$. This is so,
indeed:  $$\eqalignno{ & A(u) = \12 \left( \e^u q^{1/2}  K^{1/2} -
\e^{-u} q^{-1/2}  K^{-1/2} \right) \cr & B(u) = \12 \left( q-q^{-1}
\right) q^{-1/2}\ F\ K^{1/2} \cr
  & C(u) = \12 \left( q-q^{-1} \right) q^{-1/2}\ E\ K^{1/2} &()\cr
  & D(u) = \12 \left( \e^u q^{1/2}  K^{-1/2} -  \e^{-u} q^{-1/2}  K^{1/2}
\right) \cr }  $$ Please check that the algebraic relations \(mx232)
follow from those of the quantum group $U_q(A_1)$, equations \(mx295).

 The  affine quantum group $U_q(A_1^{(1)})$ enjoys also a  {bi-algebra}
structure, determined by the  {co-product}
$$\eqalignno{ &\Delta (E_i) = E_i \otimes K_i  + \one \otimes E_i \cr
 &\Delta (F_i) = F_i \otimes \one  +  K_i^{-1}\otimes F_i &(mx298) \cr
&\Delta (K_i) = K_i \otimes  K_i \cr }  $$
It is thus natural to look for an {intertwiner} $\cR$--matrix for the
tensor product of two spin--$\12$ irreps $(\e^{u_1}, \12) \otimes
(\e^{u_2}, \12)$ of $U_q(A_1^{(1)})$: $$\cR (\e^{u_1}, \e^{u_2})
\Delta_{\e^{u_1}, \e^{u_2}}  (g) = \Delta'_{\e^{u_1}, \e^{u_2}}  (g)
\cR(\e^{u_1}, \e^{u_2} ) \qquad \forall g \in U_q(A_1^{(1)})\tag mx299$$
This is nothing but the affinized version of the intertwiner condition
 \(mx291) for $U_q(A_1)$.  At the risk of offending the reader, we show the
 transposed co-multiplication $\Delta'$ of the generators of
$U_q(A_1^{(1)})$: $$\eqalignno{ &\Delta' (E_i) = E_i \otimes \one  + K_i
\otimes E_i \cr
 &\Delta' (F_i) = F_i \otimes  K_i^{-1}  + \one\otimes F_i &() \cr
&\Delta' (K_i) = K_i \otimes  K_i \cr } $$
Comparing \(mx298) with  \(mx278), we see that
 the relation between $E$, $F$ and $S^+$, $S^-$ is
$$   E= K^{\12} S^+ \qquad, \qquad F = K^{-\12} S^-   \tag$$

After some straightforward manipulations,   from \(mx299) we get the affine
$\cR^{\12\12}(\e^{u_1},\e^{u_2})  $ matrix: $$\eqalignno{ &
{\cR_{01}^{01} \left(\e^{u_1},\e^{u_2} \right) \over  \cR_{00}^{00}
\left(\e^{u_1},\e^{u_2}\right)} = { {\e^{u_1-u_2}}  - {\e^{u_2-u_1}} \over
q{\e^{u_1-u_2}} - q^{-1} {\e^{u_2-u_1}}} \cr  &
{\cR_{01}^{10} \left(\e^{u_1},\e^{u_2}\right) \over \cR_{00}^{00}
\left(\e^{u_1},\e^{u_2}\right)} = { q-q^{-1}  \over q{\e^{u_1-u_2}} -
q^{-1} \e^{u_2-u_1}} &(mx2100)\cr}$$   with
$\cR_{00}^{00}=\cR_{11}^{11}$,   $\cR_{01}^{01}=\cR_{10}^{10}$,
$\cR_{01}^{10}=\cR_{10}^{01}$, and all other matrix elements of $\cR$
equal to   zero.

 Identifying now $q=\e^{i\gamma}$ and $\e^u= \e^{u_1-u_2} $, we see that
$\cR(\e^{u_1}, \e^{u_2})$ is just the six--vertex $\cR $--matrix \(mx242).
With this happy result we conclude our preview  the relation between the
six--vertex model and the affine quantum group $U_q(A_1^{(1)})$.

 \subchap{Hopf algebras}
 Let $(\cA,m,\iota)$ be an algebra whose  multiplication $m:\cA\otimes
\cA\to\cA$ is associative, that is
$$[m(m\otimes \one)] (a\otimes b\otimes c) =[m(\one\otimes m)] (a\otimes
b\otimes c)\qquad \forall a,b,c\in \cA\tag$$    We write
$ab=m(a\otimes b)$, $\forall a,b\in \cA$.  If $a \in\cA$ and
 $\lambda\in\complex$, to make formal sense of $\lambda a\in \cA$ we need
the unit  map  $\iota:\complex\to\cA$ of $\cA$, which is intimately tied
to the identity     $\one\in\cA$: $$\iota:\lambda\in\complex\mapsto
\lambda\one\in\cA \tag $$  The  unit $\iota$ and the multiplication $m$ are
 compatible in the   sense that  $$ m(a\otimes \iota(\lambda)) = a\lambda
= \lambda a = m(\iota(\lambda)\otimes a) \qquad \forall a \in\cA \quad
\forall \lambda \in \complex \tag$$

 Let us consider now a   co-multiplication or  {co-product} $\Delta:
 \cA\to\cA\otimes\cA$, which should   be ``co-associative'': $$
(\Delta\otimes \one)(\Delta (a)) =  (\one\otimes\Delta )(\Delta (a))
\qquad \forall a \in \cA \eqno(coassocios)$$
 We also need a  {co-unit} map $\epsilon:\cA\to\complex$ to define a
co-algebra $(\cA, \Delta, \epsilon)$; it must satisfy
 $$ ( \one\otimes \epsilon ) \Delta = ( \epsilon \otimes \one ) \Delta =
\one \tag$$

A simultaneous algebra and {co-algebra} $(\cA,m,\iota,\Delta,\epsilon)$ is
called a  {bi-algebra} if  the  co-multiplication $\Delta$ and the co-unit
$\epsilon$  are consistent with the multiplication $m$, that is if they
are homomorphisms:  $$ \epsilon(ab)=\epsilon(a) \epsilon(b)  \qquad,\qquad
\Delta(ab)=\Delta(a) \Delta(b) \eqno()$$ Actually,  the   unit $\iota$
and   co-unit $\epsilon $ must also be compatible:
$$\iota(\epsilon(a))=\epsilon (a) \one\qquad \qquad (\forall a\in \cA)
\tag$$

A  {Hopf algebra} is a bi-algebra   enjoying an   {antipode}
$\gamma:\cA\to\cA$, which is an antihomomorphism    $$\gamma(ab) =
\gamma(b) \gamma(a)\tag antipoda$$ satisfying the  following  condition:
$$ m( \gamma\otimes \one) \Delta (a)  = m (\one\otimes \gamma) \Delta (a)
= \epsilon(a)\,\one \qquad \forall a\in \cA \tag sev67$$ This   condition
involves all the ingredients of the bi-algebra structure. Hopf algebras
are much more interesting than mere bi-algebras.

 If the multiplication $m$ is   commutative (respectively,
not commutative) , we call the algebra commutative (respectively,
non--commutative).  Just like the multiplication, the co-multiplication
may or may not be commutative. The Hopf algebra is accordingly
co-commutative or non--co-commutative. Of primary interest are those Hopf
algebras which are neither commutative nor co-commutative.

Introduce the permutation map $$\eqalign{ \sigma:\cA\otimes\cA\ & \to
\cA\otimes\cA\cr a\otimes b \ & \mapsto b\otimes a \cr}\tag $$ which
merely interchanges the order of the operands. Commutativity means thus
$$ab\equiv m(a\otimes b) = m(\sigma(a\otimes b)) \equiv m(b\otimes a)
\equiv  ba \qquad \forall a,b \in \cA \tag$$ On the other hand, if the
algebra is co-commutative, then   $$\Delta (a) = \sigma \cdot \Delta(a)
\equiv \Delta'(a) \qquad\qquad \forall a\in \cA \tag$$

 Given a co-multiplication $\Delta$, it is not hard to check that the
operation $\Delta'=\sigma \circ\Delta \in {\rm End}\,(\cA\otimes\cA)$ is
also a co-multiplication, with modified antipode $\gamma'(a) =
[\gamma(a)]^{-1}$, $(\forall a \in \cA)$. Given a Hopf algebra $\cA$, it
is called quasi-triangular if there exists a  {universal $\cR$--matrix}
$\cR\in \cA\otimes \cA$ such that  $$\Delta'(a) = \cR \Delta (a) \cR^{-1}
 \qquad \forall a\in\cA \tag m1$$ and  $$\eqalign{  (\one \otimes \Delta )
 \cR = \cR_{13} \cR_{12}  =  \sum_{i,j} A_i A_j \otimes B_j \otimes B_i
\cr  (\Delta \otimes \one ) \cR = \cR_{13} \cR_{23}  =  \sum_{i,j} A_i
\otimes A_j \otimes  B_i B_j  \cr }\tag m2$$
and  $$(\gamma\otimes \one )\cR =  (\one \otimes \gamma^{-1}) \cR =
\cR^{-1}  \tag m3$$ where $\cR$ is called the universal $\cR$--matrix. We
write $\cR=\sum_i A_i \otimes B_i$ and let  $$\eqalignno{ &\cR_{12}=\sum_i
A_i \otimes B_i \otimes \one \cr &\cR_{13}=\sum_i A_i  \otimes \one \otimes
B_i &(m4) \cr &\cR_{23}=\sum_i  \one \otimes A_i \otimes B_i  \cr} $$
Essentially,  {quasi-triangularity} means that the co-multiplication
$\Delta$ and its ``transposed'' $\Delta'$ are related linearly. In some
sense, it establishes an equivalence between two different ways
of ``adding things up''.

 A  co-commutative  algebra is trivially quasi-triangular,
with $\cR=\one\otimes \one$.   A  Hopf algebra is called  triangular  if
$\cR_{12} \cR_{21}=\one\otimes \one$, where $\cR_{21}=\sum B_i \otimes
A_i$.

 A non--co-commutative quasi-triangular  {Hopf algebra} is called a
{quantum group}.

 The interest in quasi-triangular Hopf algebras is that they produce
solutions to the Yang--Baxter equation    naturally. Indeed, from  \(m1)
and \(m2)  the Yang--Baxter equation in a ``universal form'' without
spectral parameter may be derived: $$\cR_{12} \cR_{13} \cR_{23} = \cR_{23}
\cR_{13} \cR_{12} \tag zbr$$ The proof goes like this: $$\eqalignno{
\left[ (\sigma \circ \Delta ) \otimes \one \right] \cR \ & = \sum_i
\Delta'(A_i) \otimes B_i \cr &= \sum_i \cR_{12} \Delta (A_i) \cR_{12}^{-1}
\otimes B_i \cr &=\cR_{12} \left( \sum_i \Delta(A_i) \otimes B_i \right)
\cR_{12}^{-1} &(grilloss)  \cr&= \cR_{12} \left[ (\Delta \otimes \one )
\cR \right] \cR_{12}^{-1}  \cr &= \cR_{12} \cR_{13} \cR_{23} \cR_{12}^{-1}
\cr }   $$ On the other hand,  $$\eqalignno{ \left[ (\sigma \circ \Delta )
 \otimes \one \right] \cR \ & =\sigma_{12}  (\Delta \otimes \one) \cR \cr&=
\sigma_{12} (\cR_{13} \cR_{23}) & (grilluss) \cr&=\cR_{23}  \cR_{13} \cr}
$$ and thus \(zbr) follows.

\subchap{The quantum group $U_q({\cal G})$}
In this section, we present  the    quantum semi-simple
algebras due to  Drinfeld and  Jimbo, generalizing the construction of
$U_q(s\ell(2))$ above.

 Let $\cal G$ be a  semi-simple Lie algebra with $A=(a_{ij})$
 ($i,j=1,\ldots,n= {\rm rank}\ {\cal G}$) the corresponding  {Cartan
matrix} and  $D=(D_i)$ the vector or diagonal matrix such that $D_i
a_{ij}=a_{ij} D_j $.

 The quantum group  {$U_q({\cal G})$} is defined as the algebra of formal
power series in $q$ with generators $e_i$, $f_i$, $k_i$ ($i=1,\ldots,
n={\rm rank}\ {\cal G}$) subject to the following relations:
 $$\eqalignno{ & k_i k_j = k_j k_i \cr &k_i e_j = q_i^{a_{ij}} e_j k_i \cr
&k_i f_j = q_i^{-a_{ij}} f_j k_i &(cg791) \cr
& e_i f_j - f_j e_i = \delta_{ij}\ { k_i
-k_i^{-1} \over q_i - q_i^{-1} } \cr}  $$
 $$\eqalignno{ & \sum_{\ell=0}^{1-a_{ij}} (-1)^\ell \left[ {1-a_{ij} \atop
\ell } \right]_{q_i} e_i ^{1-a_{ij} -\ell } e_j e_i^\ell = 0 \quad i\not=j
&(cg792 a)\cr & \sum_{\ell=0}^{1-a_{ij}} (-1)^\ell \left[ {1-a_{ij} \atop
\ell } \right]_{q_i} f_i ^{1-a_{ij} -\ell } f_j f_i^\ell = 0 \quad i\not=j
&(cg792 b)\cr } $$
We have used the notations
 $$  q_i = q^{D_i} \quad , \qquad  [x]_{q_i} = {q_i^x -q_i^{-x} \over q_i
 - q_i^{-1} }  \tag$$
The equations \(+11) are called the quantum  {Serre relations}. For the
simplest case of ${\cal G} = A_1$, there are no Serre relations and \(+12)
coincide with \(mx295).

The generators $e_i$, $f_i$ and $k_i$, the index $i$ ranging over the
positive simple roots,  constitute the Chevalley basis of the algebra.
Supplemented with the Serre relations, it is equivalent to the  Cartan
basis (one raising operator for each positive root).  In the quantum case,
the  {Chevalley basis} is much more convenient than the Cartan basis, due
to the profusion of $q$--factors.

The co-multiplication of $U_q({\cal G})$ is given by  $$\eqalignno{ &
\Delta (k_i ) = k_i \otimes k_i \cr & \Delta( e_i) = e_i \otimes
 k_i + \one \otimes e_i &()  \cr  & \Delta( f_i) = f_i \otimes
\one + k_i^{-1} \otimes f_i \cr } $$  and the antipode by
 $$  \gamma(e_i) = -e_i k_i^{-1} \quad, \quad \gamma(f_i) = -k_i f_i
\quad, \quad \gamma(k_i) = k_i^{-1}  \tag$$

To write the universal $\cR$--matrix of $U_q({\cal G})$ we need the
 ``logarithm'' of $k_i$, $$H_i = {1\over 2 h D_i} \log \left( k_i^2
\right)  \tag$$ where we
have set $q=\exp h  $.  Then  $${\cal R} =
 \exp \left[ h \sum_{i,j=1}^n \left( B^{-1} \right) _{ij} H_i \otimes H_j
\right]\  \left[ 1 + \sum_{i=1}^n \left( 1-q_i^{-2} \right) e_i \otimes
f_i + \cdots \right] \tag$$ where $B_{ij}=D_i a_{ij}$ is the  symmetrized
Cartan matrix.  This $\cR$--matrix follows form Drinfeld's quantum double
construction.

 In practical applications, we are interested in the $R$--matrix in some
representation. For example,   the  $R$--matrix  in the fundamental of
$U_q(s\ell(n))$, $R_n=P \cR^{({\bf n}, {\bf n})}$, is $$R_n = q
 \sum_{i=1}^{n}  e_{ii} \otimes e_{ii} + \sum_{i\not= j}  e_{ij}\otimes
 e_{ji} + \left( q  -q^{-1} \right)  \sum_{i<j} e_{jj} \otimes e_{ii}
\tag$$ where $e_{ij}$ is an $n\times n$ matrix whose only non--zero entry
is the $(i,j)$-th one.    In matrix
notation, $$\left(R_n\right)_{ij}^{k\ell} = \cases{ q  & if $i=j=k=\ell$
\cr 1 & if $i=\ell\not=k=j$\cr q-q^{-1} & if $i=k<\ell=j$ \cr 0& otherwise
\cr} \tag $$ It is not hard to verify that  $$\left(R_n\right)^{-1} =
q^{-1} \sum_{i=1}^{n}  e_{ii} \otimes e_{ii} + \sum_{i\not= j}
e_{ij}\otimes e_{ji} + \left( q^{-1} -q  \right)  \sum_{i>j} e_{jj}
\otimes e_{ii} \tag$$ and thus  $$R_n - \left(R_n\right)^{-1}
= \left( q  -q^{-1} \right) \; \one_{V\otimes V} \tag tur1$$
 where $V$ is the $\bf n$--dimensional space on which $U_q(s\ell(n))$ is
represented.

\subchap{Comments}
 In these lectures, we have tried to give a  flavor for some of the main
ideas and techniques underlying two--dimensional integrable systems. Much
has been left out, notably the thorny eight--vertex model, all the face
models (closely linked with conformal field theories), and the funny
models arising from quantum groups when $q$ is a root of unit. This last
point alone deserves a full course. Also, we have bypassed all the
applications of and to knot theory.  In a finite amount of time, however,
only so much information can be humanly absorbed.  The subject of theories
with quantum symmetries  is under active research, particularly from the
point of view of string theory: the emphasis there is  on continuum
two--dimensional field theories with $N=2$ supersymmetry.

 Continued and passionate discussions with Mat\'{\i}as Moreno are
gratefully acknowledged.  I wish to thank the continued strenuous efforts
of Jos\'e Luis Lucio, the organizer and soul of the School (now ten years
old), for succeeding in creating every two years a convivial atmosphere
where scientific discussion can be carried out freely and invigoratingly.
My thanks also to this year's most worried organizer, Miguel Vargas. I am
much indebted to the interesting questions and inquisitive minds of the
participants at the School.   Last but not least, I am thankful to the
Instituto de F\'{\i}sica de la Universidad Nacional Aut\'onoma de M\'exico
(IFUNAM) for bestowing on me the C\'atedra Tom\'as Brody 1992. This work
 has been partially supported by the Fonds National Suisse pour la
Recherche Scientifique.

\subchap{References}
 It is perhaps more useful to point out a few good review articles and
books;   references to  the original literature  can be found easily
starting from the works cited below.

The book on which these notes are based:

 \br  C. G\'omez, M. Ruiz--Altaba and G. Sierra|1993|Quantum Groups in
Two--Dimensional    Physics|Cambridge University Press|Cambridge|

Nice introductions to the theory of factorizable $S$--matrices:

\jr    A.B. Zamolodchikov and   Al.B.
 Zamolodchikov|1979|Factorized $S$--matrices in two
dimensions as the exact solution of certain relativistic
quantum field theory models|Ann. Phys.|120|253|

\jr   A.B. Zamolodchikov|1980|Factorized $S$--matrices in lattice
statistical systems|Sov. Sci. Rev.|A2|1|

The Bethe ansatz:

 \bibitem{J.H. Lowenstein ($1982$). Introduction to Bethe ansatz approach
in ($1+1$)--dimensional models, in {\it Proceedings,  Les Houches XXXIX}
(Elsevier, Amsterdam).}

 \jr L.D. Faddeev and L.A. Takhtajan|1981|What is the spin of a
 spin wave?|Phys. Lett.|85A|375|

Vertex models:

\br  R.J.  Baxter|1982|Exactly Solved Models in
Statistical Mechanics|Academic Press|London|

The Leningrad school:

\vr E.K. Sklyanin|1993|Quantum inverse scattering method:
selected topics|Proceedings of
 the Fifth Nankai Workshop|Mo-Lin Ge|World Scientific|Singapore|

Quantum groups:

\vr V.G. Drinfeld|1987|Quantum groups|Proceedings of the 1986
International Congress of
Mathematics|A.M.~Gleason|Am. Math. Soc.|Berkeley|

\br M. Jimbo|1991|Quantum Groups and the Yang--Baxter
Equation|Springer|Tokyo|

\jr L.D. Faddeev, N.Yu. Reshetikhin and
 L.A. Takhtajan|1990|Quantum Lie
groups and Lie algebras|Leningrad Math.~J.|1|193--225|

\jr L. Alvarez--Gaum\'e, C. G\'omez and G.
Sierra|1989|Hidden
quantum symmetries in rational conformal field theories|Nucl.
 Phys.|B319|155|

 \jr V. Pasquier and H. Saleur|1990|Common structures between finite
systems and conformal field theories through quantum groups|Nucl.
Phys.|B330|523|

\end